\documentclass[prb,aps]{revtex4}

\usepackage{graphicx}
\usepackage{epsfig}
\usepackage{dcolumn}
\usepackage{bm}

\begin{document}

\title{Classical dynamics of a nano-mechanical resonator coupled to a
single-electron transistor}

\author{A. D. Armour}\affiliation{School of Physics and
Astronomy, University of Nottingham, Nottingham NG7 2RD, United
Kingdom}
\author{M. P. Blencowe and Y. Zhang} \affiliation{Department of Physics and Astronomy,
Dartmouth College,
Hanover, NH 03755}

\date{\today}

\begin{abstract}

We analyze the dynamics of a nano-mechanical resonator coupled to
a single-electron transistor (SET) in the regime where the
resonator behaves classically.  A master equation is derived
describing the dynamics of the coupled system which is then used
to obtain equations of motion for the average charge state of the
SET and the average position of the resonator. We show that the
action of the SET on the resonator is very similar to that of a
thermal bath, as it leads to a steady-state
probability-distribution for the resonator which can be described
by mean values of the resonator position, a renormalized
frequency, an effective temperature and an intrinsic damping
constant.  Including the effects of extrinsic damping and finite
temperature, we find that there remain experimentally accessible
regimes where the intrinsic damping of the resonator still
dominates its behavior. We also obtain the average current through
the SET as a function of the coupling to the resonator.

\end{abstract}


\maketitle

\section{Introduction}
Recent developments in fabrication have made it possible to
produce nano-electromechanical systems in which mechanical
resonators with frequencies in the range 100 MHz--1 GHz are
coupled electrostatically to mesoscopic
conductors.\cite{Henry,roukes,blickrev,elect,QD,QPC,set} The
dynamics of nano-electromechanical systems have attracted
considerable interest because of the novel transport mechanisms
they can give rise to,\cite{cs,shutrev,qs,nish} their interesting
non-linear and chaotic properties\cite{nonlin,blickrev,ron} and
their applications as fast and ultra-sensitive
sensors.\cite{roukes,elect} Furthermore, when cooled to ultra-low
temperatures, high-frequency resonators are expected to display
quantum mechanical behavior and so nano-electromechanical devices
could be used to explore the cross-over from quantum to classical
behavior in mechanical systems.\cite{ABS,wp,squeeze,noise}

Theoretical,\cite{white,bw,ZB} and experimental,\cite{set} work
has suggested that a single-electron transistor (SET) could form
the basis of an extremely sensitive motion detector for
nano-mechanical resonators. Therefore, the question of what the
potential sensitivity of such a displacement detector will be, and
how it is determined by the underlying dynamics of the coupled
SET--resonator system, is of practical as well as theoretical
interest.

When a mechanical resonator, coated with a thin metallic layer, is
placed next to a SET, it acts like a capacitor whose presence
affects the current flowing through the SET. The capacitance, and
therefore the current, depend on the position and hence the motion
of the resonator. The sensitivity with which the displacement of a
resonator can be measured from the current through the SET is
limited by two sources of noise. Firstly, the shot noise in the
current through the SET limits its sensitivity. Secondly, the
fluctuating charge on the SET island will act as a fluctuating
force on the resonator, resulting ultimately in displacement
noise. Thus far, the sensitivity of the displacement detection
which can be achieved has been gauged by treating the resonator
and the SET as two distinct elements.\cite{ZB} Although this
approach provides a reasonably accurate estimate of the
sensitivity, it neglects correlations between the displacement
noise of the resonator and the shot noise in the current. In fact,
the coupled SET and resonator form an intrinsically interesting
dynamical system in their own right. By studying the full coupled
dynamics of the system a broad understanding of the underlying
physics can be achieved which can then be applied to obtain
important insights into when the system would function most
sensitively as a measuring device.

In this article, we analyze the coupled dynamics of the
SET--resonator system, treating the resonator as a classical
harmonic oscillator. In particular, we examine the effect on the
motion of the resonator of electrons tunnelling through the SET
and the concomitant effects of the resonator motion on the average
current of the SET. More generally, our analysis of the coupled
dynamics provides useful insights into the dynamical behavior of
nano-electromechanical systems.

The dynamics of a {\it quantum} harmonic oscillator coupled to an
electrical tunnel junction was studied recently by Mozyrsky and
Martin\cite{noise} and by Smirnov {\it et al.}\cite{noise2} In the
regime where the drain-source voltage across the junction, $V_{\rm ds}$,
is much larger than the energy quanta of the oscillator,
$\hbar\omega_0$, Mozyrsky and Martin found that the dynamics of
the density matrix of the oscillator reduced to a form very
similar to that of the Caldeira--Leggett model\cite{CL} for an
oscillator in contact with a thermal bath. From the connection
with the Caldeira--Leggett model, Mozyrsky and Martin were able to
show that the tunnelling electrons caused rapid dephasing of the
oscillator, acting like a thermal bath with an effective
temperature and associated damping constant.

For  a classical resonator coupled only to a SET,
we find that the electrons affect the resonator in a similar way
to the quantum system considered by Mozyrsky and
Martin.\cite{noise} For sufficiently weak coupling between the
resonator and the SET, the electrons act like a thermal bath which
heats the resonator to an effective temperature. However, if the
coupling between the SET and the resonator is increased the
behavior of the resonator begins to deviate significantly from
that of an oscillator in contact with a thermal bath. Including
the effects of extrinsic damping of the resonator, we find that
the effective temperature of the resonator is suppressed.

The classical regime for the resonator which we consider here
will be the appropriate description for systems where $eV_{\rm
ds}\gg \hbar\omega_0$ and the model we use  to treat
the electron tunnelling processes in the SET will require $eV_{\rm
ds}\gg k_{\rm B}T_{e}$,\cite{ult} where $T_{e}$ is the electron temperature
(assumed to be the same in the SET island and leads). As the electrons passing
through the SET  heat the resonator to an
effective temperature which is proportional to $eV_{\rm ds}$ in the
absence of extrinsic damping, this means that the resonator effective
temperature can be much larger than the electron temperature. We will
assume possible overheating of the SET can be
neglected,\cite{martinis,korotemp}
so that this temperature imbalance can be maintained and the above
model assumptions continue to hold.

 The outline of this paper is as follows. In
Sec.\ II, we review the properties of the SET with a fixed voltage
gate and the master equation method which can be used to determine
the average current. Then in Sec.\ III, we generalize the results
for a fixed voltage gate to the case where the gate is a
nano-mechanical resonator. Initially, we analyze the dynamics of
the resonator coupled to the SET ignoring external influences. We
derive the equations of motion for the probability distribution of
the system and show that, with certain approximations, the average
properties of the system can be described by simple, physically
intuitive equations of motion. We investigate the steady-state
properties of the probability distribution for the coupled system
by numerical integration of the master equations that govern the
dynamics.

In Sec.\ IV, we consider the effect of damping in the resonator,
arising from sources other than the interaction with the SET, and
the effect of finite background temperature. We examine the
current characteristics of the system in Sec.\ V, and obtain the
average current through the SET when it is coupled to the
resonator. Then in Sec.\ VI, we discuss how our results could be
investigated experimentally. Finally, in Sec. VII, we draw our
conclusions and discuss the more general implications of our
results for understanding the dynamics of nano-electromechanical
systems.

In the appendices we give further details about aspects of our
calculations. In appendix A, we derive the Hamiltonians and
electron tunnelling rates for the coupled SET--resonator system.
We give details of the method used to integrate the master
equations for the SET--resonator probability distribution in
appendix B. The effect of higher vibrational modes of the
nano-mechanical resonator is discussed in appendix C.  Finally, in
appendix D we estimate the typical experimentally achievable
values of the parameters appearing in our model.

\section{Charge dynamics of a SET coupled to a fixed gate}
The development of the single-electron transistor
(SET)\cite{set_nat} and its subsequent refinement\cite{rfset}
offers the prospect of carrying out electrometry with
unprecedented accuracy.\cite{elect2} A SET consists of a small
metal island joined to leads by two tunnel junctions. The small
capacitance of the metal island can give rise to a Coulomb
blockade where the charging energy of a single electron dominates
over thermal effects at low temperatures and the only accessible
state of the system has a fixed number of electrons on the island,
$N$. An external voltage-gate can be used to induce a polarization
charge on the metal island which makes the states with $N$ and
$N+1$ electrons degenerate, and thus allows current to flow
through the device. In the low temperature regime, states with
more than $N+1$ or fewer than $N$ electrons have much higher
energies because of charging effects and so play no role in the
dynamics of the system.\cite{FG}

Before we go on to consider the coupled dynamics of the
SET-resonator system, we briefly review the properties of the SET
when coupled to a fixed voltage-gate. The circuit diagram for a
nano-mechanical resonator coupled to the island of a
single-electron transistor is shown schematically in Fig.\ 1.
Neglecting the effect of the dynamics, the resonator acts as a
simple voltage-gate with a fixed position. Indeed, the dynamics of
the resonator, which we consider in the next section, generally
acts as a small perturbation on top of the effect of a static
voltage gate.

\begin{figure}[t]
\center{ \epsfig{file=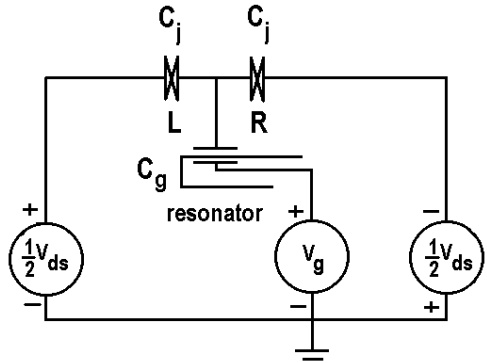, width=8.0cm}} \caption{Circuit
diagram for the coupled SET-resonator system. The SET island lies
between the two tunnel junctions which are assumed to have the
same capacitance $C_j$. Under the drain-source voltage shown, the
electrons tend to flow through the SET island from right to left.
}
\label{fig:schema}
\end{figure}

We simplify the analysis a little by assuming that the SET is
symmetrical, as shown in Fig.\ 1. The two junctions linking the
metal island to the leads have the same capacitances, $C_j$,
resistances, $R$, and a drain-source voltage $V_{\rm ds}$ (which we take
to be positive throughout) is applied across the island
symmetrically. The static resonator acts as the gate of a
capacitor, $C_g$, with a voltage $V_g$ applied to it. The  gate
voltage can have a strong influence on the charge dynamics of the
electrons on the island because of the polarization charge it
induces.

The energy of $N$ charges on the SET island with a fixed gate is
given, up to an unimportant constant factor,  by\cite{Makhlin}
\begin{equation}
H_N=\frac{e^2}{2C_{\Sigma}}(N^2-2NN_g),
\end{equation}
where $C_{\Sigma}=2C_j+C_g$ is the total  capacitance of the SET,
and $N_g=-C_gV_g/e$ is the polarization charge induced by the
gate.

In the regime where $e^2/(2C_{\Sigma})\gg k_{\rm B}T_{e}$, and
where the bias voltage is not too large, the electron tunnelling
processes in the SET are restricted to those between $N$ and $N+1$
electrons on the island.\cite{bw,ZB,Makhlin} The tunnelling
dynamics of the electrons can be modelled by a simple master
equation known as the orthodox model.\cite{FG} This approach is
valid provided that the effective tunnel junction resistances
satisfy $R>h/e^2$, the charging energy dominates over the thermal
energy, and the SET is biased well away from the Coulomb blockade
regions so that co-tunnelling can be neglected.

We now outline the form of the master equation and the way in
which the steady-state current is obtained for the SET with a
fixed gate. The analysis of the coupled dynamics of the
SET-resonator system is then carried out by  generalizing these
results to include the motion of the resonator.

\subsection{Master Equation}

Within the orthodox model,\cite{FG} the state of an ensemble of
SETs for which only the two charge states $N$ and $N+1$ are
accessible can be described by the probabilities $P_N(t)$ and
$P_{N+1}(t)$, of finding $N$ or $N+1$ electrons on the island at
time $t$, respectively. These probabilities evolve in time
according to a pair of coupled master equations,
\begin{eqnarray}
\frac{dP_N}{dt}&=& -(\Gamma_L^{-}+\Gamma_R^{+})P_N+(\Gamma_L^{+}+
\Gamma_R^{-})P_{N+1} \\
\frac{dP_{N+1}}{dt}&=&
(\Gamma_L^{-}+\Gamma_R^{+})P_N-(\Gamma_L^{+}+\Gamma_R^{-})P_{N+1},
\end{eqnarray}
where $\Gamma_{L(R)}^{\pm}$ are the tunnelling rates forwards
($+$) or backwards ($-$) across the two junctions $L$ and $R$
between the island and the leads, as illustrated in Fig.\
\ref{fig:schema}. Here we use the convention that the forward
direction is from right to left, matching the direction set by the
drain-source voltage. The tunnelling rates are calculated using
the orthodox model and at finite temperature, $T_{e}$, are given
by\cite{FG}
\begin{equation}
\Gamma_{L(R)}^{\pm}=\frac{1}{e^2R}\frac{\Delta
E_{L(R)}^{\pm}}{1-{\rm e}^{-\Delta E_{L(R)}^{\pm}/k_{\rm B}T_{e}}},
\label{ft}
\end{equation}
where $\Delta E_{L(R)}^{\pm}$ is the initial free energy (of the
SET and leads) before tunnelling through the $L(R)$ junction,
minus the final free energy after tunnelling forwards ($+$) or
backwards ($-$). At $T_{e}=0$, this expression reduces to the
simplified form,
\begin{equation}
\Gamma^{\pm}=\frac{1}{e^2R}\Delta E_{L(R)}^{\pm}\Theta(\Delta
E_{L(R)}^{\pm}),
\end{equation}
where $\Theta(x)$ is the Heaviside step-function.

\subsection{SET Current}

 The current through the SET depends on the
tunnelling rates through the two junctions. In the steady-state,
current conservation means that we need only ever consider
electron tunnelling processes across one junction. When the
Coulomb-blockade limits the accessible charge states of the SET
island to $N$ and $N+1$, the steady-state current can be written
as\cite{ford}
\begin{eqnarray}
\langle I \rangle&=& e\left(\Gamma_{L}^+ P_{N+1} -\Gamma_{L}^-P_N\right)\\
&=&e\left(\Gamma_{R}^+P_{N} -\Gamma_{R}^-P_{N+1}\right),
\end{eqnarray}
where $\langle \cdots \rangle$ denotes an ensemble average and where
${P}_{N}$ and ${P}_{N+1}=1-{P}_N$
are the probabilities of having $N$ and $N+1$ electrons on
the island, respectively. Substituting in the correct expressions for the
tunnelling rates, we find
\begin{equation}
\langle I\rangle=\frac{e}{\tau_t}{P}_N{P}_{N+1},
\label{currfix}
\end{equation}
where  $\tau_t=Re/V_{\rm ds}$ is
a measure of the average time between tunnelling events.

\section{Dynamics of the Coupled SET and Resonator}
Unlike a fixed voltage-gate, a movable voltage-gate, such as a
nano-mechanical resonator, responds dynamically to the force
exerted on it by electrons tunnelling through the SET island.
Furthermore, the motion of the resonator acts back on the SET by
altering the tunnelling rates of the electrons. The coupled
dynamics of the SET electrons and the resonator eventually results
in a steady-state, even without including the effects of external
sources of dissipation on the resonator.

The dynamics of the SET-resonator system are described by
generalizing the master equation for the SET charges to include
the position and velocity of the resonator. Crucially, the
tunnelling rates for electrons in the master equation must be
re-calculated to include the effect of the resonator's motion.
Although the master equation for the SET-resonator system is
apparently much more complicated than that for the fixed-gate
case, the dynamics of the average resonator position and SET
charge are readily determined once certain simplifying
approximations are made. Furthermore, the master equation can be
solved numerically and it is found that for sufficiently weak
coupling the steady-state probability distributions are
approximately Gaussian in form. The effect of the SET on the
motion of the resonator can then be gauged very accurately from
the first and second moments of the steady-state probability
distribution.

We treat the nano-mechanical resonator as a single-mode harmonic
oscillator, as it turns out that the interaction between the SET
and the fundamental flexural mode of the resonator dominates the
behavior. However, if we assume that interactions between the
various vibrational modes of the nano-mechanical resonator can be
neglected, then the results we obtain for a single mode can be
generalized to include higher frequency modes.

\subsection{Master Equation}
The probability distribution for the SET-resonator system is a
function of the oscillator position, $x$, and velocity, $u$, as
well as the charge state of the SET island. We can characterize
the state of the system by the probability distribution pair,
$P_{N}(x,u;t)$ and $P_{N+1}(x,u;t)$, which give the probabilities
of the oscillator being at position $x$, with velocity $u$, and
having $N$ or $N+1$ electrons on the island, respectively, at time
$t$. At $T_{e}=0$, the equations of motion for these probability
distributions take the form,
\begin{eqnarray}
\frac{dP_N(x,u;t)}{dt}&=&\left\{H_N,P_N(x,u;t)\right\}\\
\nonumber
 &&+\frac{1}{Re^2}\left[ \Theta(\Delta E_L^{+})\Delta
E_L^{+}+
 \Theta(\Delta E_R^{-})\Delta E_R^{-}\right]P_{N+1}(x,u;t)\\
 \nonumber
&&-\frac{1}{Re^2}\left[ \Theta(\Delta E_R^{+})\Delta
E_{R}^{+}+\Theta(\Delta E_L^{-})\Delta E_L^{-}\right]P_N(x,u;t)\\
\label{pn}
\frac{dP_{N+1}(x,u;t)}{dt}&=&\left\{H_{N+1},P_{N+1}(x,u;t)\right\}\\
\nonumber &&+\frac{1}{Re^2} \left[ \Theta(\Delta E_R^{+})\Delta
E_{R}^{+} +\Theta(\Delta E_L^{-})\Delta
E_L^{-}\right]P_{N}(x,u;t)\\ \nonumber &&-\frac{1}{Re^2}\left[
\Theta(\Delta E_L^{+})\Delta E_L^{+}+ \Theta(\Delta E_R^{-})\Delta
E_R^{-}\right]P_{N+1}(x,u;t), \label{pn1}
\end{eqnarray}
where $H_{N(N+1)}$ is the Hamiltonian for the SET-resonator system
with $N(N+1)$ electrons on the island, and $\{\cdot,\cdot\}$ is a
Poisson bracket. The quantities $\Delta E^{\pm}_{L(R)}$ are the
total free energy differences for an electron tunnelling forwards
($+$) or backwards ($-$) across the left-hand (right-hand)
junction.

The Hamiltonians for the coupled SET-resonator system with $N$ and
$N+1$ electrons on the SET island are readily obtained treating
the resonator as a single-mode harmonic oscillator, with angular
frequency $\omega_0$, centered a distance $d$ away from the SET
island. Details of the calculation are given in appendix A. The
Hamiltonians take the form
\begin{eqnarray}
H_N&=&E_c\delta N +\frac{p^2}{2m}+\frac{1}{2}m\omega_0^2x^2\\
H_{N+1}&=&-E_c\delta N
+\frac{p^2}{2m}+\frac{1}{2}m\omega_0^2(x-x_0)^2-\frac{1}{2}m\omega_0^2x_0^2\\
&&-m\omega_0^2x^2_0(N_g-\delta N -1/2 ),
\end{eqnarray}
where $\delta N=N_g -N-1/2$ and $x_0=-2E_cN_g/(m\omega_0^2 d)$.
The length-scale $x_0$ has a simple physical interpretation: it is
the distance between the equilibrium positions for the resonator
with $N$ and $N+1$ electrons on the SET island.

The energy differences which determine the tunnelling rates are
given by
\begin{eqnarray}
\Delta E^+_L&=&-\Delta E^-_L= -2E_c\delta
N-m\omega_0^2x_0x-m\omega_0^2x^2_0(N_g-\delta N -1/2 )+\frac{eV_{\rm ds}}{2}\\
\Delta E^+_R&=& -\Delta E^-_R=2E_c\delta
N+m\omega_0^2x_0x+m\omega_0^2x^2_0(N_g-\delta N -1/2
)+\frac{eV_{\rm ds}}{2},
\end{eqnarray}
as derived in appendix A. Notice that at $T_{e}=0$, for a given
displacement of the resonator and voltage bias, tunnelling across
each junction will only be able to occur in one direction. The
equations of motion for the probability distributions can be
extended to finite temperatures by replacing the tunnelling rates
by the appropriate finite temperature form, as given in Eq.\
(\ref{ft}).

\subsection{Mean-Coordinate Dynamics}

Integration of the coupled master equations gives a complete
description of the ensemble dynamics. However, the
equations of motion for the probability distribution are
complicated and it is not easy to extract an overall picture of
the dynamics from them. Nevertheless, by making a few physically
motivated approximations, it is possible to obtain equations of
motion for all of the first and second moments of the
distributions.

In order to simplify the analysis, we assume that $\delta N$,
$V_{\rm ds}$, and the range of $x$ for which the probabilities,
$P_N(x,u;t)$ and $P_{N+1}(x,u;t)$, are non-negligible are such
that $\Delta E_L^+,\Delta E_R^+$ are always positive. Such an
approximation is valid so long as the drain-source voltage is the dominant
energy scale so that $eV_{\rm ds}\gg m\omega_0^2 x_0^2, E_C\delta
N$ and the distributions $P_N(x,u;t)$ and $P_{N+1}(x,u;t)$ are
strongly peaked at $x\sim x_0$.

Taking into account this approximation, we need only consider two
out of the four tunnelling processes, as the step function
constraints will always be satisfied for the processes in the
`$+$' direction, and never satisfied for processes in the `$-$'
direction. Thus, we obtain a simplified pair of coupled master
equations,
\begin{eqnarray}
\frac{dP_N}{dt}&=&\omega_0^2x\frac{\partial P_N}{\partial
u}-u\frac{\partial P_N}{\partial x}+\frac{1}{Re^2}\left(E_L
P_{N+1}-E_R P_N-m\omega_0^2 x_0 x P\right )\label{spn} \\
\frac{dP_{N+1}}{dt}&=&\omega_0^2(x-x_0)\frac{\partial
P_{N+1}}{\partial u}-u\frac{\partial P_{N+1}}{\partial
x}-\frac{1}{Re^2}\left(E_L P_{N+1}-E_R P_N-m\omega_0^2 x_0
xP\right), \label{spn1}
\end{eqnarray}
where $P(x,u;t)=P_N(x,u;t)+P_{N+1}(x,u;t)$, and we have defined
$E_L=-2E_c\delta N-m\omega_0^2x^2_0(N_g-\delta N -1/2 )+{eV_{\rm
ds}}/{2}$ and $E_R=2E_c\delta N+m\omega_0^2x^2_0(N_g-\delta N -1/2
)+{eV_{\rm ds}}/{2}$ (i.e., $E_L$ and $E_R$ are the
position-independent parts of $\Delta E^+_L$ and $\Delta E^+_R$,
respectively).

Adding together Eqs.\ (\ref{spn}) and (\ref{spn1}), the tunnelling
terms cancel out and we obtain the simplified equation of motion
for the full probability distribution.
\begin{equation}
\frac{dP}{dt}=\omega_0^2x\frac{\partial P}{\partial
u}-u\frac{\partial P}{\partial x}-x_0\frac{\partial
P_{N+1}}{\partial u}. \label{tp}
\end{equation}
We can now obtain equations which describe the dynamics of the
ensemble-averaged position and velocity by multiplying Eq.\ (\ref{tp}) by $x$
and $u$ respectively and then integrating over both $x$ and $u$,
\begin{eqnarray}
\langle \dot{u}(t)\rangle &=& -\omega_0^2\langle x(t)\rangle
+\omega_0^2 x_0 \langle P\rangle_{N+1}(t)
\label{mfu}\\
 \langle \dot{x}(t)\rangle &=& \langle u(t)\rangle \label{mfx}
\end{eqnarray}
where the averages are defined by
\[
\langle \cdots \rangle=\int dx\int
du\:(\cdots)(P_{N+1}(x,u;t)+P_N(x,u;t))
\]
and
\[
\langle P\rangle_{N+1}=\int dx \int du\:P_{N+1}(x,u;t).
\]
We can obtain a closed set of equations by integrating Eq.\
(\ref{spn1}) over $x$ and $u$ to obtain,
\begin{equation}
Re^2\dot{\langle P\rangle}_{N+1}(t)=E_R-eV_{\rm
ds}\langle P\rangle_{N+1}(t)-m\omega_0^2x_0\langle x(t) \rangle.
\label{mfp}
\end{equation}

\subsubsection{Fixed Point Analysis}

The steady-state behavior of the system is described by the
fixed-points of the mean-coordinate equations
(\ref{mfu},\ref{mfx},\ref{mfp}). Setting
$\langle\dot{u}\rangle=\langle\dot{x}\rangle=\dot{\langle
P\rangle}_{N+1}=0$, we find that there is a single solution at
\begin{eqnarray}
\langle{u}\rangle_{\rm fp}&=&0\\
\langle {x}\rangle_{\rm fp}&=&x_0(\langle P\rangle_{N+1})_{\rm fp} \label{xfp}\\
(\langle P\rangle_{N+1})_{\rm fp}&=&\frac{2E_C\delta
N+m\omega_0^2x^2_0(N_g-\delta N -1/2 )+eV_{\rm ds}/2}{eV_{\rm
ds}-m\omega_0^2 x_0^2}.
\end{eqnarray}

In order to make the physical content of the mean-coordinate
equations (\ref{mfu},\ref{mfx},\ref{mfp}) more apparent, we
transform them to variables centered on their fixed points,
\begin{eqnarray}
\langle \tilde{x} \rangle&=&\langle x \rangle
-\langle x\rangle_{\rm fp}\\
\langle\tilde{P}\rangle_{N+1}&=&\langle P\rangle_{N+1}-
(\langle P\rangle_{N+1})_{\rm fp},
\end{eqnarray}
and we also re-express them in terms of the dimensionless
frequency and coupling strength parameters $\epsilon=\omega_0 \tau_t$ and
$\kappa=m\omega_0^2x_0^2/(eV_{\rm ds})$, respectively. Hence we find,
\begin{eqnarray}
\langle\dot{\tilde{P}}\rangle_{N+1}&=&\frac{\kappa}{x_0}\langle\tilde{x}
\rangle-\langle{\tilde{P}}\rangle_{N+1}\\
\langle\dot{\tilde{x}}\rangle&=&\langle{{u}} \rangle\\
 \langle\dot{{u}}\rangle&=&\epsilon^2(x_0\langle{\tilde{P}\rangle}_{N+1}-
 \langle{\tilde{x}}\rangle).
\end{eqnarray}
where the time coordinate has been scaled by the tunnelling time,
$\tau_t$.

 The scaled
parameter $\epsilon$ compares the relative time-scales of the
oscillator period and electron tunnelling time. Generally, the
oscillator motion will be much slower than the electron tunnelling
time and so we expect $\epsilon\ll 1$ for most practical
situations. The parameter $\kappa$ is a measure of the interaction
strength between the SET and the resonator in terms of the
energy-scale defined by the drain-source voltage. In practice, the
coupling between the resonator and the SET which can be achieved
will be weak so that $\kappa\ll 1$, as discussed in Sec.\ VI.

 The mean-coordinate dynamics about the fixed point is given
by a set of three coupled, linear, first-order differential
equations, and the stability of the fixed point can be determined
from the behavior of the associated eigenvalues. Although the
eigenvalues of a system of three equations can be determined
analytically, the expressions obtained are algebraically rather
complicated. However, the stability of the system is determined by
whether or not the real part of any of the eigenvalues becomes
positive, in which case the system can become unstable. A simple
numerical calculation shows that the SET-resonator system is
entirely stable for $\kappa<1$, but for $\kappa>1$ it seems that
our equations predict that the system becomes unstable.
However, any potential instability in the motion of the resonator
would clearly involve processes which lie outside the domain of
our approximate description and so we cannot draw any conclusions
about the overall stability of the system within the existing
model.

\subsubsection{Analytical Description}

The set of linear equations describing the mean-coordinate
dynamics of the SET-resonator system can be solved analytically,
and the general solution obtained. However, as with the
eigenvalues, the general solution is algebraically too complicated
to be useful in developing an intuitive picture of the dynamics.
However, in the regime where $\epsilon\ll 1$ and $\kappa<1$ a very
good approximation to the full solution can be derived as an
expansion in $\epsilon$.

\begin{figure}[t]
\center{ \epsfig{file=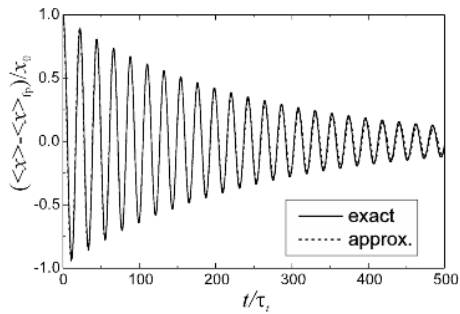, width=6.0cm}}
\caption{Comparison of the approximate analytical equations of
motion for the resonator with the exact equations, integrated
numerically. In this plot $\kappa=0.1$ and $\epsilon=0.3$, so the
approximate equations are close to the limit on their validity
($\epsilon\ll 1$).} \label{fig:compare}
\end{figure}

The approximate solutions for the trajectory of the system take
the form,
\begin{eqnarray}
\langle \tilde{x}(t)\rangle &=&{\rm e}^{-\kappa\epsilon^2
t/2\tau_t}\left[ \tilde{x}(0)\cos\left(\sqrt{1-\kappa}~\epsilon t
/\tau_t\right)
+\frac{u(0)}{\sqrt{1-\kappa}~\epsilon}\sin\left(\sqrt{1-\kappa}~
\epsilon t/\tau_t\right)\right] \label{dho1}\\
 \langle\tilde{P}\rangle_{N+1}(t) &=&\left(\tilde{P}_{N+1}(0)-
 \kappa\frac{\tilde{x}(0)}{x_0}\right){\rm
 e}^{-(1-\kappa\epsilon^2)t/\tau_t} \nonumber \\
 &+&{\rm
 e}^{-\kappa\epsilon^2t/2\tau_t}\frac{\kappa}{x_0}\left[\tilde{x}(0)
 \cos\left(\sqrt{1-\kappa}~
 \epsilon t/\tau_t\right)
  +\frac{u(0)}{\sqrt{1-\kappa}~\epsilon}\sin
 \left(\sqrt{1-\kappa}~\epsilon t/\tau_t\right)\right] \label{dho2}
\end{eqnarray}
with the initial conditions
$\langle\tilde{x}(t=0)\rangle=\tilde{x}(0)$, $\langle
u(t=0)\rangle=u(0)$ and
$\langle\tilde{P}_{N+1}\rangle(t=0)=\tilde{P}_{N+1}(0)$. The
approximate trajectory is in fact very close to the full solution,
as is shown in Fig.\ \ref{fig:compare} for $\kappa=0.1$ and
$\epsilon=0.3$. The value of $\epsilon$ in the plot is at the
limit of the range of validity of the approximate equations, but
the agreement remains good. For values of $\epsilon<0.1$ the
approximate and exact curves overlay each other so well that the
two curves cannot easily be distinguished.

The approximate analytical solution for the dynamics of the
oscillator given by Eq.\ (\ref{dho1}) can be obtained from the
equation of motion for a damped harmonic oscillator, with damping
constant $\kappa\epsilon^2/\tau_t$, and (angular) frequency
$\sqrt{1-\kappa}~\epsilon/\tau_t$. The shift in frequency is due
to the interaction between the oscillator and the tunnelling
electrons.\cite{CL} We can think of $\gamma_{i}=\kappa
\epsilon^2/\tau_t$ as an {\it intrinsic} damping rate for the
SET-resonator system, to be distinguished from {\it extrinsic}
damping arising from sources other than interaction with the SET
(see subsection~\ref{extrinsic}).

\subsection{Numerical Solution of the Coupled Master Equations}

As well as extracting mean-coordinate properties, it is possible
to integrate the equations for the probability distributions
[Eqs.\ (\ref{spn}) and (\ref{spn1})] numerically, using a method
which is outlined in appendix B. Numerical integration allows us
to demonstrate that the coupled master-equations do indeed lead to
a steady-state probability distribution and also to explore how
this steady-state is reached.

\begin{figure}[t]
\center{\epsfig{file=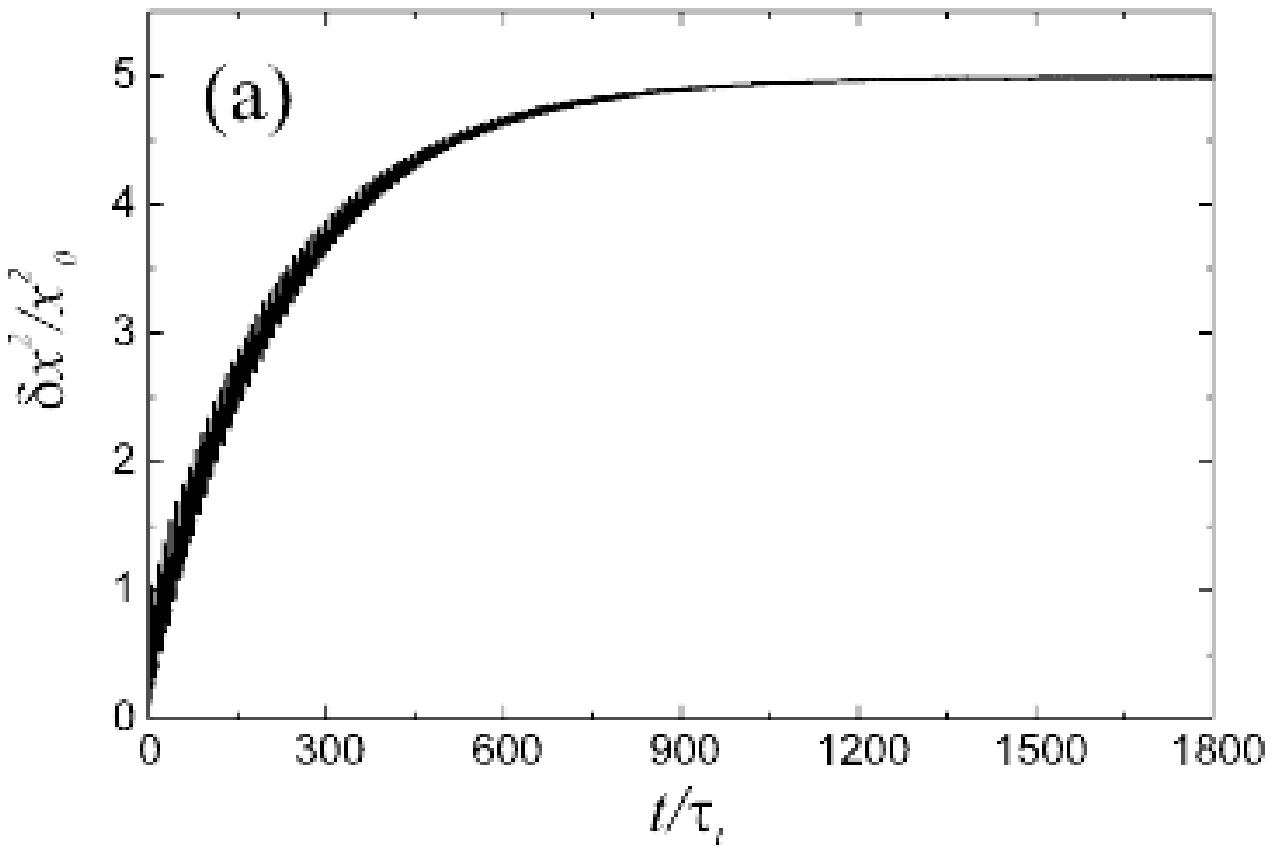, width=6.0cm}
 \epsfig{file=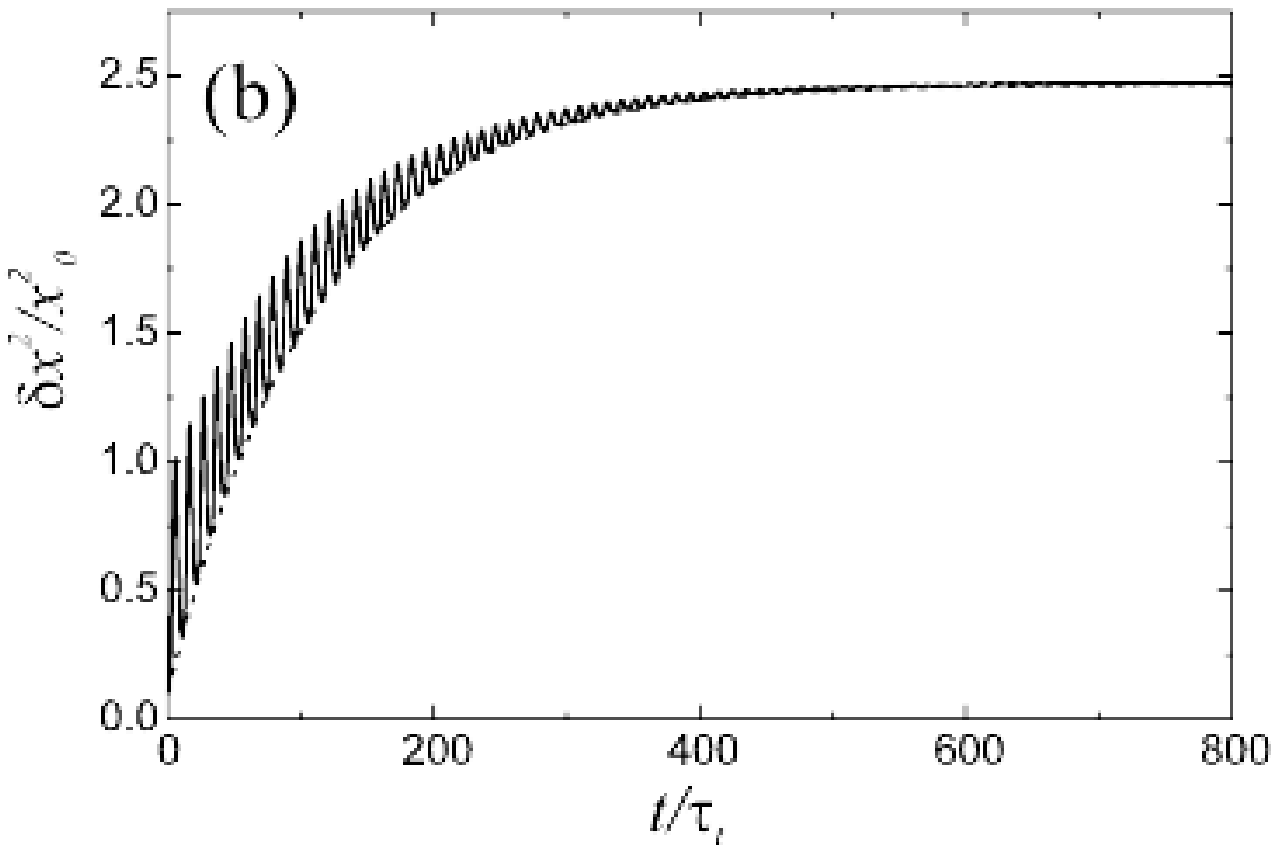,width=6.0cm}
 \epsfig{file=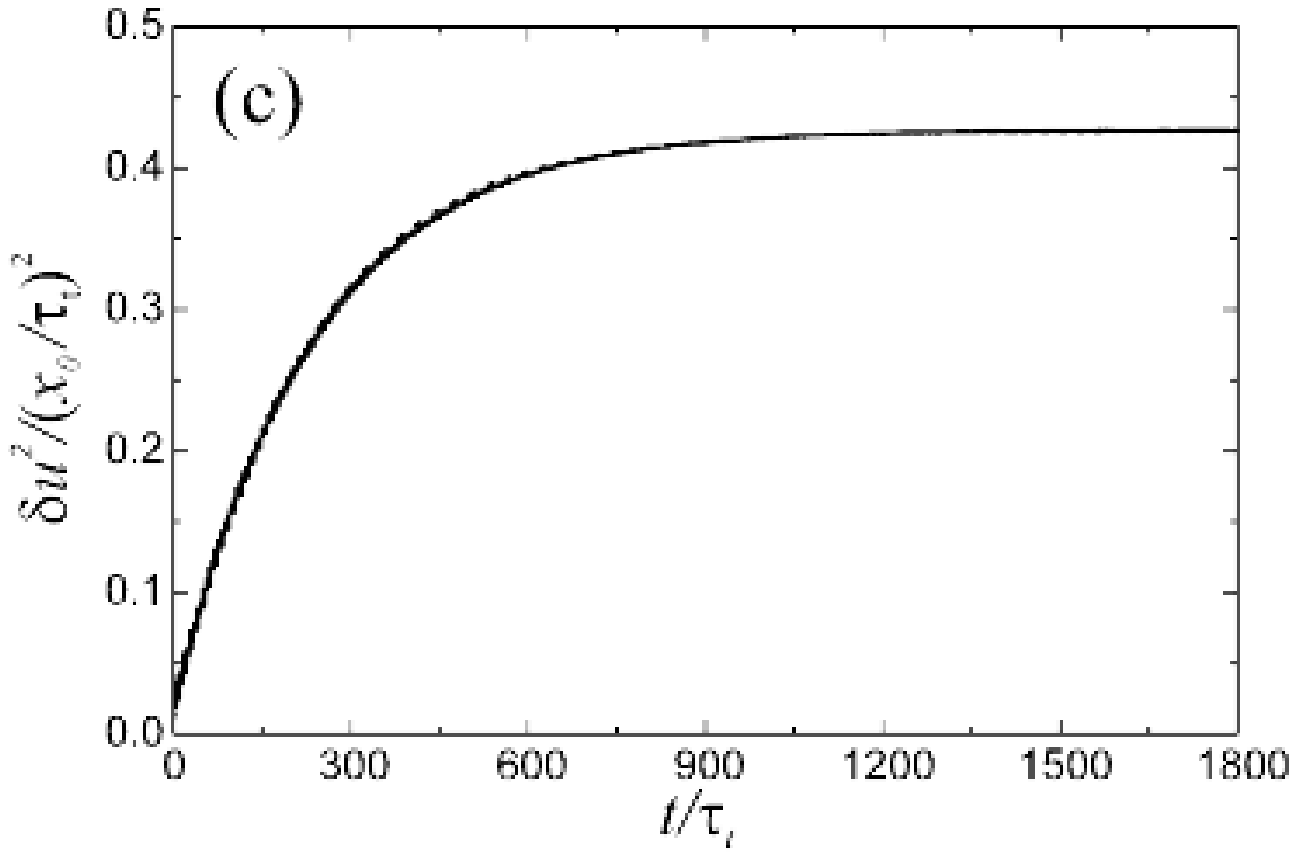,width=6.0cm}
  \epsfig{file=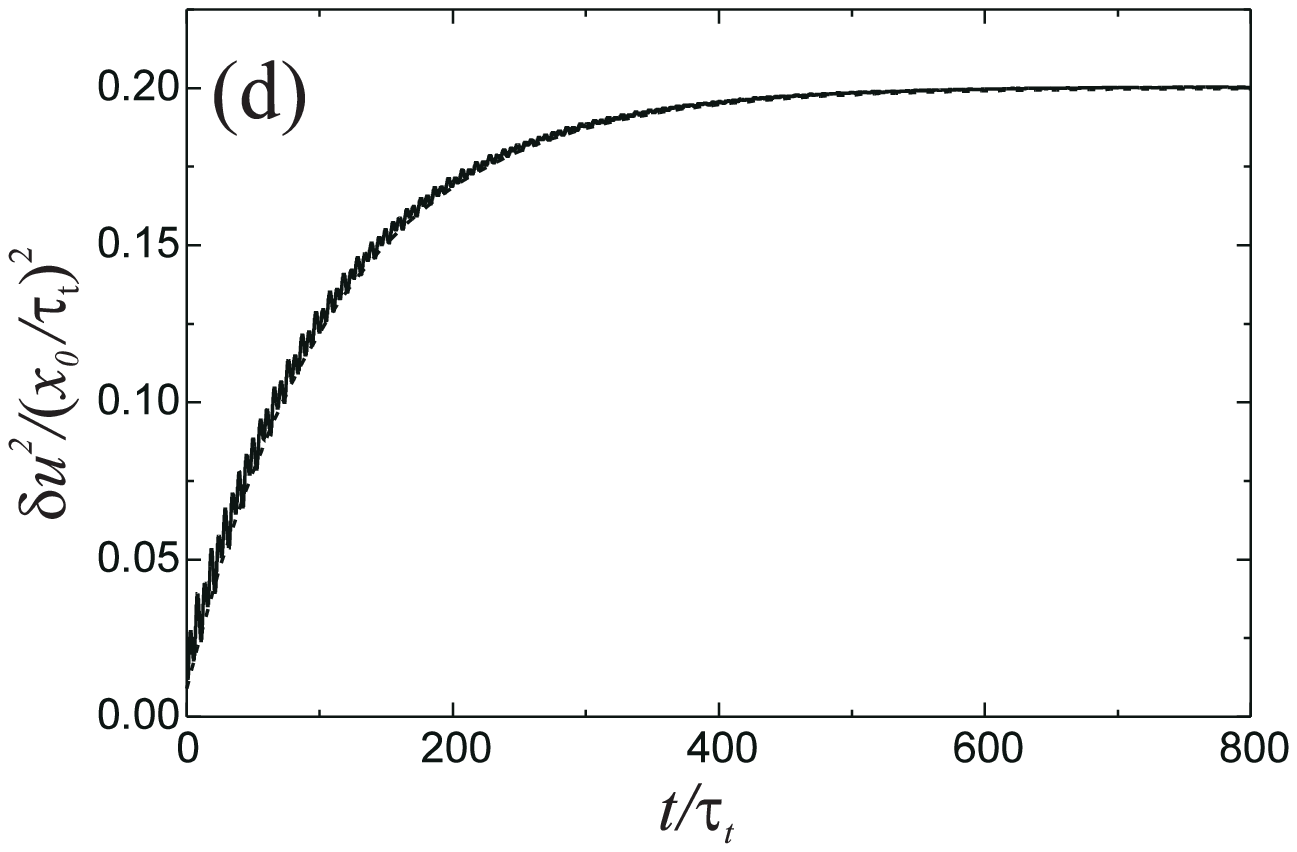,width=6.0cm}}
  \caption{Behavior of the variances, $\delta x^2$ and
$\delta u^2$, as the steady-state is approached. In panels (a) and
(c) $\kappa=0.05$, and in panels (b) and (d) $\kappa=0.1$, with
 $\epsilon=0.3$ throughout. The dashed lines represent fits to an
 exponential time-dependence obtained using the initial variances, the
 steady-state values
 calculated using Eqs.\ (\ref{deltax}) and (\ref{deltau}), and the appropriate
 decay rate, $\gamma_{i}=\epsilon^2\kappa$, in each case.} \label{fig:one}
\end{figure}

 Figure \ref{fig:one} shows the dynamics of the variance, $\delta
 x^2$, of the full probability distribution, $P(x,u;t)$. The initial probability
 distributions were chosen to be sharply peaked in phase space. For simplicity we
 have performed all the calculations at a gate-voltage such that
 $E_L=E_R=eV_{\rm ds}/2$, so that the $N$ and $N+1$ charge states would be
 degenerate in the limit $\kappa=0$.   After
 some initial oscillations, $\delta x^2$ settles down to a steady
 value.The rate at which the steady-state
 is achieved is $\gamma_{i}$, just as we would expect for a damped
 harmonic oscillator. Notice, however, that all the lengths
 have been scaled by $x_0$, which itself depends on $\kappa$.
 Thus, the absolute magnitude of the variances increase with
 $\kappa$ in contrast to the scaled values which actually decrease.

\begin{figure}[t]
\center{ \epsfig{file=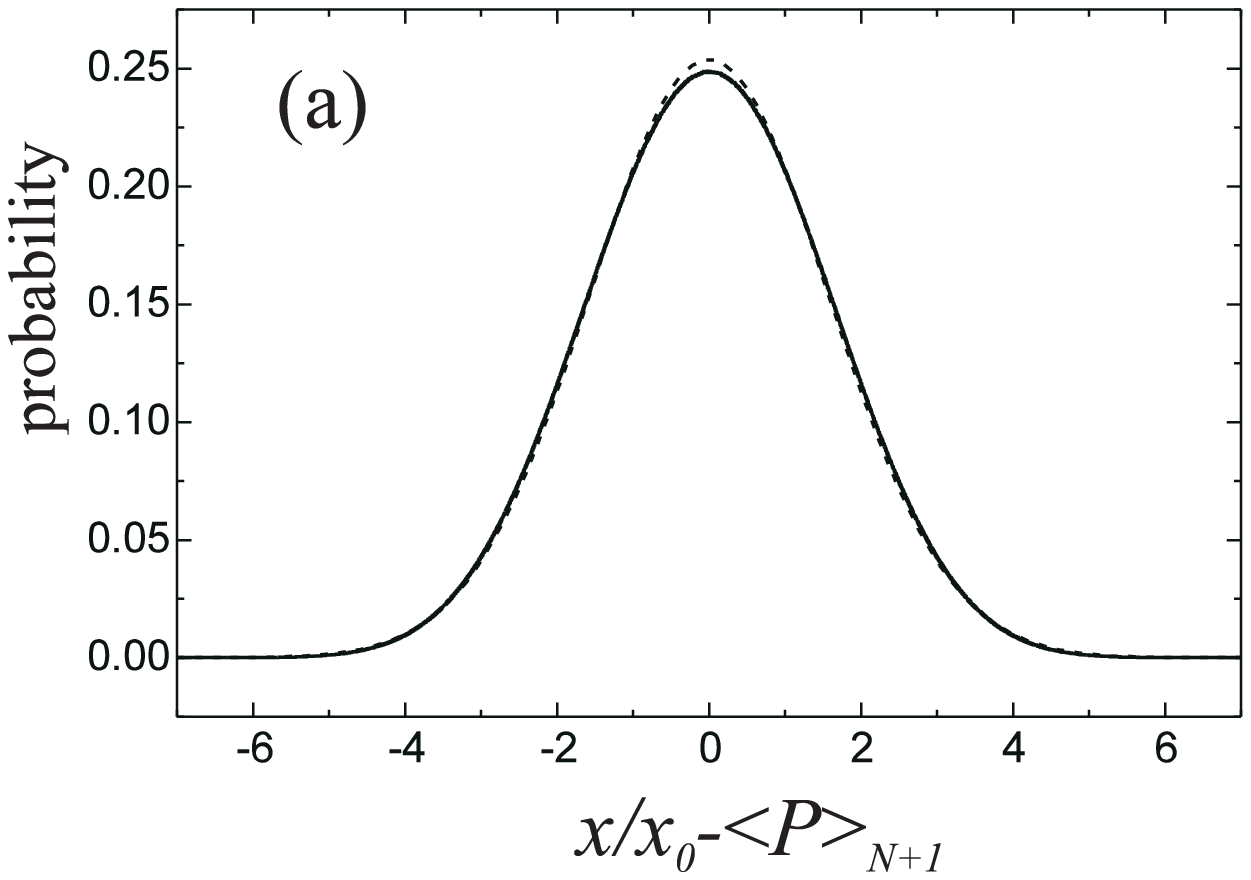, width=6.0cm}
\epsfig{file=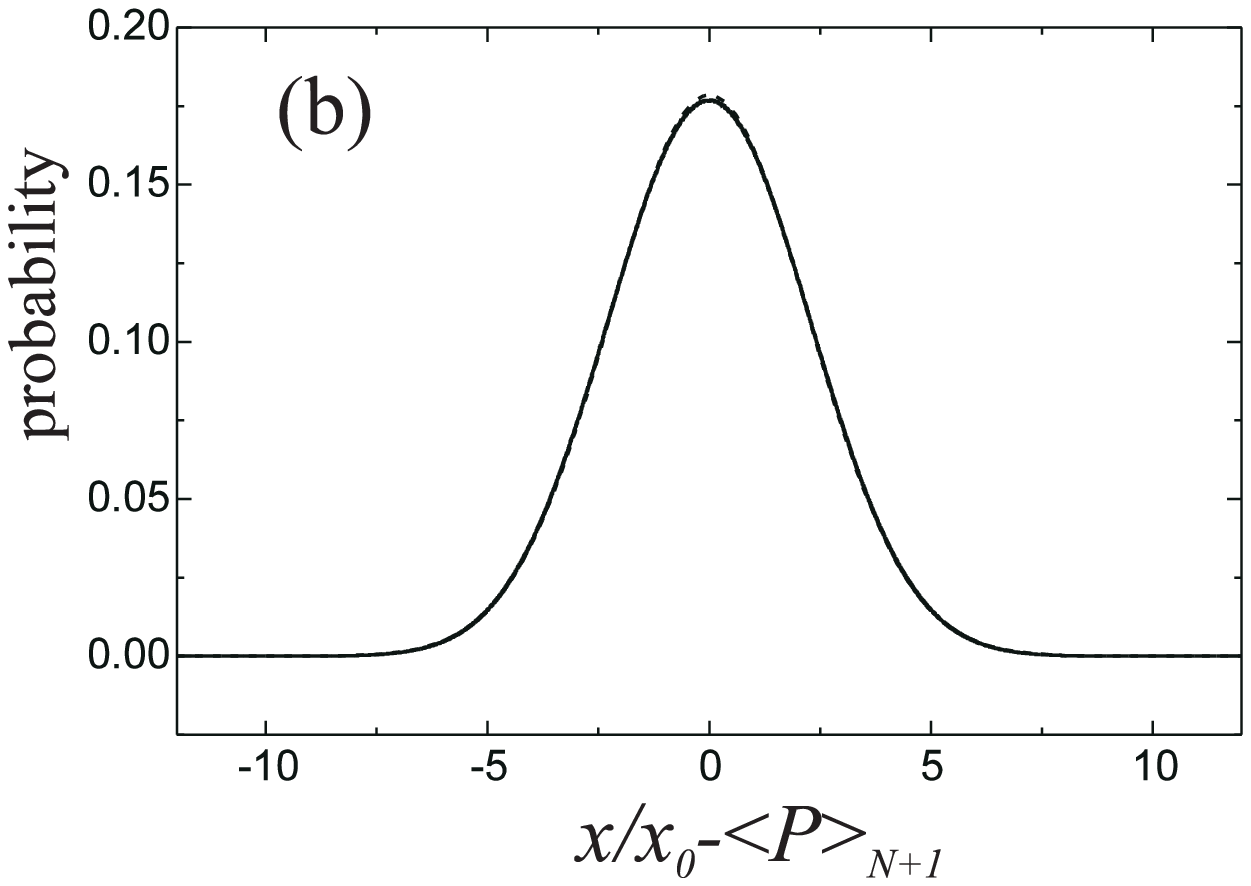,width=6.0cm}
\epsfig{file=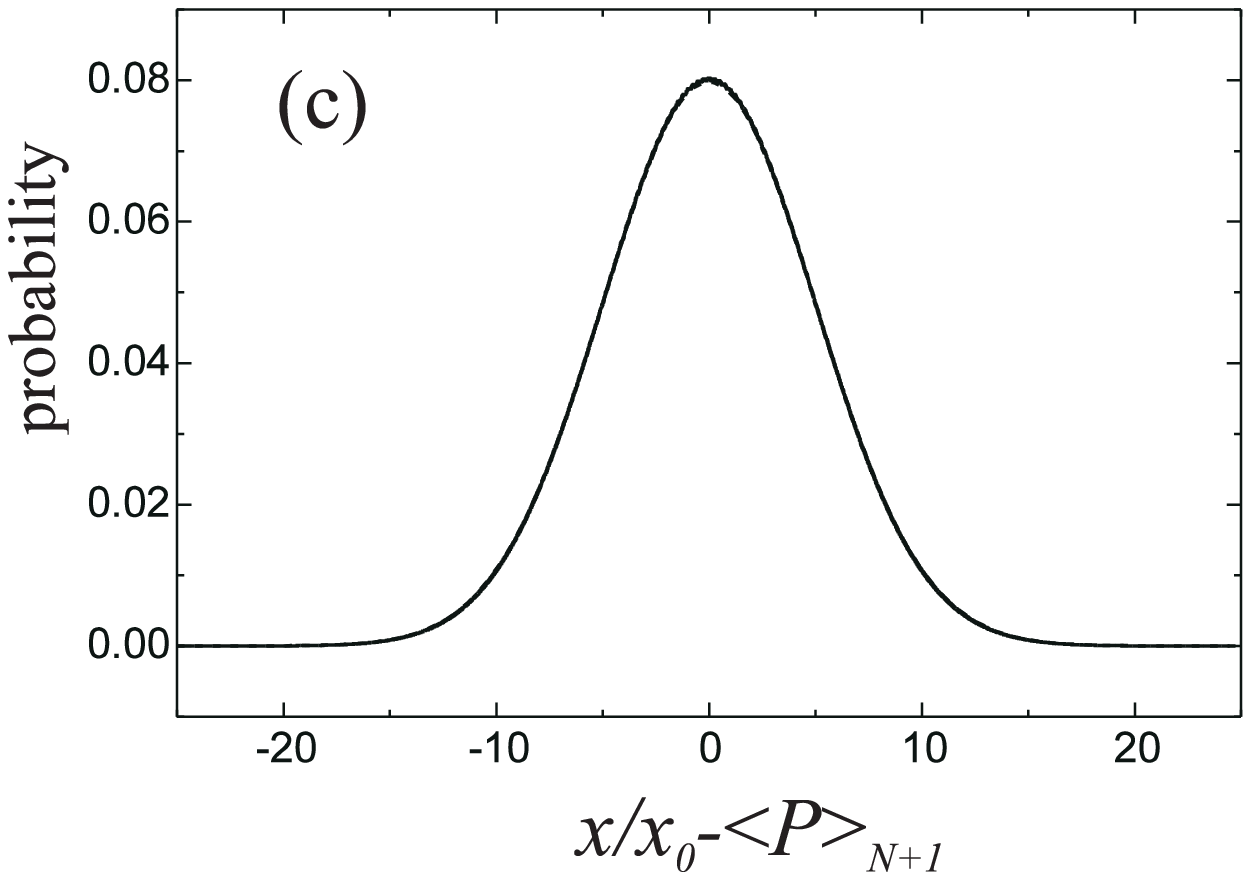,width=6.0cm} \epsfig{file=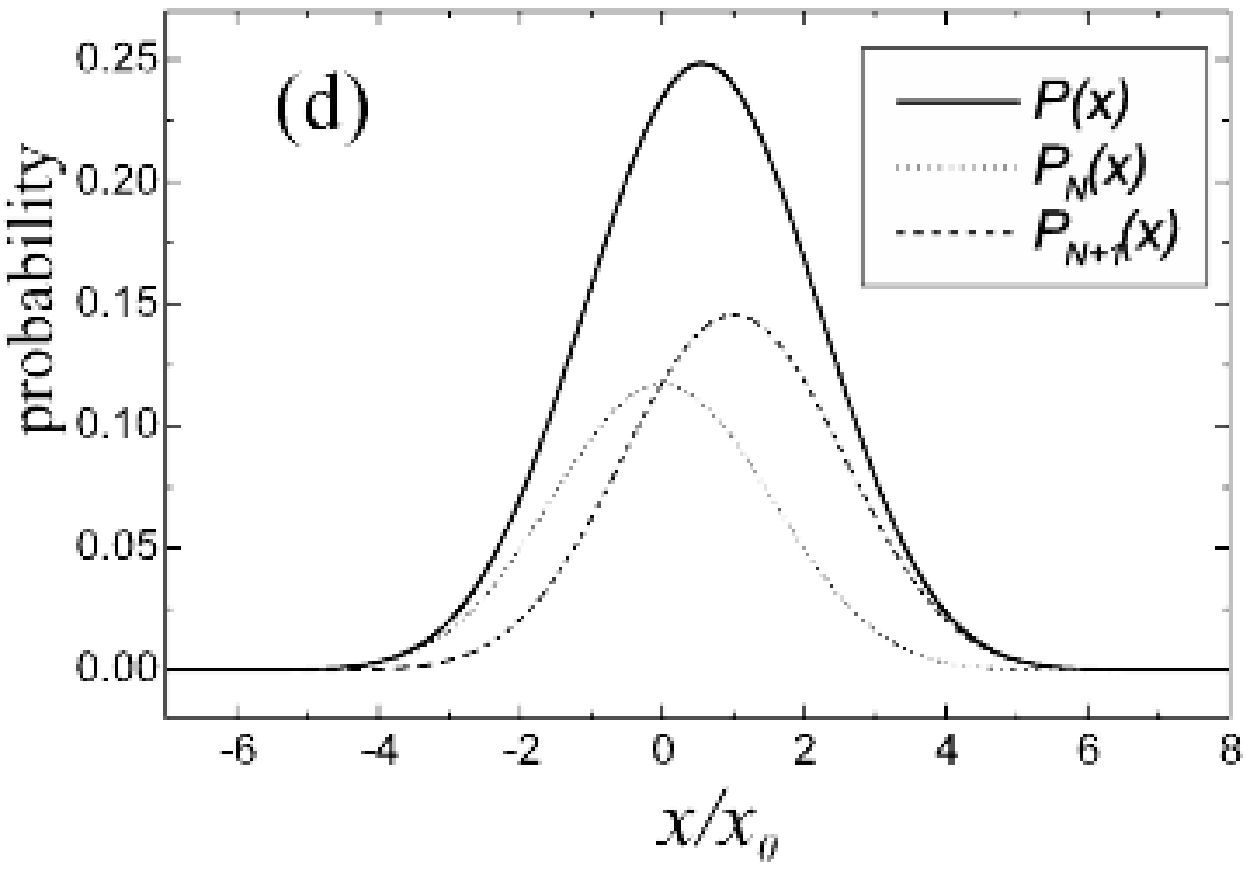,
width=6.0cm}} \caption{Steady-state probability distributions as a
function of position. The full probability distribution, $P(x)$,
centered on the fixed point $x_0\langle P\rangle_{N+1}$, is shown in
(a) with $\kappa=0.1$, (b) with $\kappa=0.05$, and (c) with $\kappa=0.01$,
where $\epsilon=0.3$ in each case. Each of the numerical curves is
compared with a Gaussian fitted using the fixed-point position and
the position variance, obtained from Eqs.\ (\ref{xfp}) and
(\ref{deltax}) respectively. In (d) the sub-distributions $P_N(x)$
and $P_{N+1}(x)$ for $\kappa=0.1$ and $\epsilon=0.3$ are shown.}
\label{fig:two}
\end{figure}

 The steady-state
probability distributions $P(x)$, for $\epsilon=0.3$ and
$\kappa=0.1, 0.05$, and $0.01$, respectively, are shown  in
comparison with the appropriate Gaussian fits in Fig.\
\ref{fig:two}. The steady-state
distributions are indeed strongly peaked, as we assumed earlier,
and the overall distribution is very nearly Gaussian as can be
seen from the plots which are generated from the average position
and associated variance in each case. The deviation from a
Gaussian distribution is most apparent for $\kappa=0.1$ and the
agreement improves as the coupling is reduced.
The sub-distributions $P_N(x)$ and $P_{N+1}(x)$ for
$\kappa=0.1$ are shown in Fig.\ \ref{fig:two}(d).
They are clearly
centered on different points and also closely
resemble Gaussians individually. However, since one cannot add two
displaced Gaussians to obtain a single Gaussian curve, it is clear
that the curves are not exactly Gaussian in form.

\subsection{Variances of the distribution}

Since the steady-state distribution is close to Gaussian for
$\kappa \ll 1$, the state of the resonator in this regime will be
almost completely specified by the average values of its
coordinates and their variances. Making use of the fact that there
is indeed a well-defined steady-state, we can derive analytical
expressions for the variances in the resonator position and
velocity.

 Multiplying  Eqs.\ (\ref{spn},\ref{spn1}) by the
appropriate quantity and integrating over the position and
velocity of the resonator, we  obtain the following closed set of
equations of motion:
\begin{eqnarray}
\langle \dot{xu}\rangle &=& -\omega_0^2\langle x^2\rangle+\langle
u^2\rangle+\omega_0^2 x_0\langle x\rangle_{N+1}\\
\langle \dot{x^2}\rangle &=& 2\langle xu \rangle\\
\langle \dot{u^2}\rangle &=&-2\omega_0^2\langle xu\rangle+2x_0\omega_0^2
\langle u\rangle_{N+1} \\
\langle \dot{x}\rangle_N &=& \langle u\rangle_N+\frac{1}{Re^2}\left[E_L\langle
x\rangle_{N+1}-E_R\langle x\rangle_N-m\omega_0^2x_0\langle x^2\rangle\right]\\
\langle \dot{x}\rangle_{N+1} &=& \langle
u\rangle_{N+1}-\frac{1}{Re^2}\left[E_L\langle
x\rangle_{N+1}-E_R\langle x\rangle_N-m\omega_0^2x_0\langle
x^2\rangle\right]\\
\langle \dot{u}\rangle_{N} &=& -\omega_0^2\langle
x\rangle_{N}+\frac{1}{Re^2}\left[E_L\langle
u\rangle_{N+1}-E_R\langle u\rangle_N-m\omega_0^2x_0\langle
ux\rangle\right]\\
\langle \dot{u}\rangle_{N+1} &=& -\omega_0^2\langle
x\rangle_{N+1}+\omega_0^2x_0\langle P\rangle_{N+1}-\frac{1}{Re^2}\left[E_L
\langle
u\rangle_{N+1}-E_R\langle u\rangle_N-m\omega_0^2x_0\langle
ux\rangle\right],
\end{eqnarray}
where $\langle \cdots \rangle_N$ and $\langle \cdots
\rangle_{N+1}$ imply averages over the sub-distributions
$P_N(x,u)$ and $P_{N+1}(x,u)$, respectively.

In the steady-state, all time derivatives will be zero and hence
we can infer the following properties of the stationary
distribution
\begin{eqnarray}
\langle x\rangle_{N+1}&=&x_0\langle P\rangle_{N+1}\\
\langle u^2\rangle-\omega_0^2\langle
x^2\rangle&=&-\omega_0^2x_0\langle x\rangle_{N+1}\\
\langle x^2 \rangle&=&\frac{E_L}{m\omega_0^2}\langle P\rangle_{N+1},
\end{eqnarray}
and $\langle xu \rangle = \langle u \rangle_N= \langle x
\rangle_N=\langle u \rangle_{N+1}=0$.
 Taken together with the steady-state solutions of the
mean-coordinate equations, $\langle
x\rangle=x_0\langle P\rangle_{N+1}$ and $\langle u\rangle=0$, we
obtain expressions for the variances in position and velocity for
the stationary probability distribution. Thus we find,
\begin{eqnarray}
\delta x^2 &=&\frac{x_0^2}{\kappa}\langle P\rangle_{N}\langle P\rangle_{N+1}=
\frac{eV_{\rm ds}}{m\omega_0^2}\langle P\rangle_{N}\langle P\rangle_{N+1}
\label{deltax}\\
 \delta u^2 &=&\frac{x_0^2\omega_0^2}{\kappa}(1-\kappa)
\langle P\rangle_{N}\langle P\rangle_{N+1}= \frac{eV_{\rm
ds}}{m}(1-\kappa)\langle P\rangle_{N}\langle P\rangle_{N+1}.
\label{deltau}
\end{eqnarray}

\begin{figure}[t]
\center{\epsfig{file=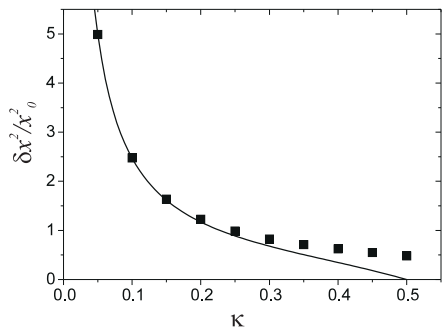, width=6.0cm} }
  \caption{Comparison of the variance, $\delta x^2$,
 calculated using Eq.\ (\ref{deltax}) (curve), and the results of a numerical
 integration
 (points). Both calculations are performed for $E_L=E_R$.} \label{fig:compare2}
\end{figure}

The accuracy of these expressions, based on the assumption that
only forward electron tunnelling occurs, can be tested by
comparing them with the results obtained from numerical
integration of the master equation. Fig. \ref{fig:compare2}
compares the analytical result, Eq.\ (\ref{deltax}), with the
numerical result as a function of $\kappa$. It is clear that
there is excellent agreement for $\kappa\leq 0.1$, but for larger
values of $\kappa$ the validity of the analytical expression
breaks down.

In order to understand the expressions for the variances, Eqs.\
(\ref{deltax}) and (\ref{deltau}), we can compare them with those
for a damped harmonic resonator in contact with a heat
bath.\cite{lemon} For an oscillator in contact with a bath at
temperature $T$, we have equipartition of energy:
$m\omega_0^2\delta x^2=m\delta u^2=k_{\rm B}T$.
Comparison with Eqs.\ (\ref{deltax}) and (\ref{deltau}) shows that
for $\kappa\ll 1$ the steady-state probability distribution for
the resonator is very close to a thermal one, and we can identify
an effective temperature, $T_{\rm eff}$, by analogy with the
thermal case,
\begin{equation}
k_{\rm B}T_{\rm eff}=eV_{\rm ds}\langle P\rangle_N\langle P\rangle_{N+1}.
\label{te}
\end{equation}

The electrons passing through the SET give rise to a fluctuating
force on the oscillator as its equilibrium position is shifted
back and forth by $x_0$, but because the oscillator acts back on
the motion of the electrons through the tunnelling rate, the
motion of the oscillator is also damped and a steady-state is
achieved. For $\kappa\ll 1$, the steady-state is very close to a
thermal state with an effective temperature which depends on
properties of the SET alone: the drain-source voltage applied
across it and the average probability of finding $N$ or $N+1$
electrons on the island. In particular, the effective temperature
does not seem to depend at all on the strength of the coupling
between the resonator and the SET. In fact, the strength of the
coupling $\kappa$ does affect the effective temperature, as the
values of $\langle P\rangle_{N+1}$ and $\langle P\rangle_{N}$
depend (albeit weakly) on $\kappa$ for $\kappa \ll 1$. However,
the value of the damping constant, $\gamma_{i}$, is proportional
to $\kappa$, and so the coupling strength controls how long it
takes for the steady state to be reached. If the coupling between
the SET and the resonator is switched off (i.e.,\ $\kappa=0$),
then Eqs.\ (\ref{deltax}) and (\ref{deltau}) break down because a
steady-state for the resonator is never achieved as there is no
damping.

It is interesting to compare the expressions for the effective
temperature and damping constant of the resonator with results for
a quantum oscillator coupled to a single electrical tunnel
junction (TJ).\cite{noise,noise2} Our results for a classical
oscillator can be compared directly with calculations for a
quantum oscillator when $eV_{\rm ds}\gg \hbar \omega_0$ so that
the electrons cause  rapid dephasing of the resonator and induce
the transition from quantum to classical behavior.\cite{noise} The
effective temperature we obtain for a resonator coupled to a SET
[Eq.\ (\ref{te})] is very similar to that obtained for a resonator
coupled to a TJ, $k_{\rm B}T_{\rm eff}=eV_{\rm ds}/2$ (with the
electron temperature set to zero in the electrodes). However,
because $\langle P\rangle_N\langle P\rangle_{N+1}\leq 1/4$, the
heating induced by the SET will always be at least a factor of 1/2
less than that induced by the TJ. In fact, the  effective
temperatures obtained for both the TJ and SET can be commonly
equated to one-half the ensemble-averaged energy dissipated by an
electron due to tunnelling across a junction. For the TJ, the
energy dissipated (for zero electron temperature in the
electrodes) is always $eV_{\rm ds}$. For the SET, on the other
hand, the average energy dissipated by a tunnelling electron is
$E_{R}\langle P\rangle_{N}+E_{L}\langle P\rangle_{N+1}=
2E_{L}\langle P\rangle_{N+1}\simeq 2eV_{\rm ds}\langle
P\rangle_N\langle P\rangle_{N+1}$ for $\kappa\ll 1$.

The intrinsic damping constant for the TJ takes the form
$\gamma_{\rm TJ}=\mu C^2/m$, to lowest order in the coupling
constant $C$, where $\mu$ is a constant which depends on the
densities of states in the leads. In contrast, for the SET the
intrinsic damping rate, $\gamma_{\rm i}=(Re^3/md^2)(V_gC_g/v_{\rm
ds}C_{\Sigma})^2$, depends explicitly on the drain-source voltage.
The difference in damping rates relates directly to the difference
in electron tunnelling rates for a TJ and SET: for the TJ the
electron tunnelling rates across the junction do not depend on the
drain-source voltage, to lowest order, whereas for the SET the
tunnelling rates always depend on the drain-source voltage in the
orthodox regime.

\section{Resonator with Extrinsic Damping and at Finite Temperature}

Thus far, we have considered the nano-mechanical resonator and the
SET as a closed system without including any external influences.
Under these circumstances, and for sufficiently weak coupling to
the resonator, the SET acts like a thermal reservoir which heats the
oscillator and also gives rise to
damping. The effective temperature in the steady-state is not a
function of the coupling strength between the SET and the
resonator, $\kappa$, but the time taken to reach that steady-state
does depend on the intrinsic damping rate which in turn depends on
$\kappa$. However, the resonator will have a finite
quality-factor, even in the absence of the SET,\cite{clelbook} and
if the value of the associated damping rate, $\gamma_{e}$, is
much larger than that due to the SET, then it will be essential to
include these damping processes in the dynamics. Furthermore, both
the resonator and the electrons will be at finite temperatures,
set by their surroundings (which need not necessarily be the
same). The effects of extrinsic damping and finite background
temperature can be taken into account by modifying the master
equation formalism we have developed so far.

\subsection{Variances with Extrinsic Damping}
\label{extrinsic}

The equation of motion for the probability distribution is readily
modified to include extrinsic damping of a harmonic
oscillator arising from sources other than its interaction with the
SET.  We can obtain the correct form for the coupled master
equations by adding the term
\[
\frac{\partial }{\partial u}\left(\gamma_e u P_{N(N+1)}  \right)
\]
to Eqs.\ (\ref{spn}) and (\ref{spn1}), where the extrinsic damping
is modelled as a force on the oscillator of the form $-\gamma_{e}
u$.\cite{lemon}

The mean-coordinate equations~(\ref{mfu},\ref{mfx},\ref{mfp}) and
the position of the fixed-point corresponding to the steady-state
are unaffected by the extrinsic damping. However, this apparent
lack of change belies a significant increase in the complexity of
the underlying dynamics. With finite extrinsic damping, the
steady-state values of $\langle x\rangle_N$ and $\langle
x\rangle_{N+1}$ start to converge so that $\langle x\rangle_N>0$
and $\langle x\rangle_{N+1}<x_0 \langle P\rangle_{N+1}$.
Furthermore, with extrinsic damping although the overall average
velocity remains zero in the steady-state, the average velocities
for given charge states are non-zero: $ \langle
u\rangle_N=-\langle u\rangle_{N+1}\neq 0$.

The variances of the steady-state distribution including extrinsic
damping are obtained using the same method as before, and we find
\begin{eqnarray}
\delta x^2&=&\frac{\delta x^2_{i}(1+\alpha)}{1+\frac{\alpha}{\epsilon^2 \kappa}
\left[(1-\kappa)(1+\alpha)+\epsilon^2\right]} \label{deltaxe}\\
 \delta u^2&=&\frac{\delta u^2_{i}}{1+\frac{\alpha}{\epsilon^2 \kappa}
\left[(1-\kappa)(1+\alpha)+\epsilon^2\right]}, \label{deltaue}
\end{eqnarray}
where $\alpha=\gamma_{e}\tau_t$, and $\delta x^2_{i}$
and $\delta u^2_{i}$ are the intrinsic variances (no extrinsic
damping) given by Eqs.\ (\ref{deltax})
and (\ref{deltau}), respectively.

In the regime where  $\kappa,\epsilon,\alpha\ll 1$, the effect of
the SET on the resonator will depend very sensitively on the ratio
$\gamma_{e}/\gamma_{i}=\alpha/(\epsilon^2\kappa)$. When
$\gamma_{e}\ll \gamma_{i}$ the extrinsic damping will not
cause any serious modification of the resonator's behavior. In
contrast, when $\gamma_{e}\gg \gamma_{i}$ the extrinsic
damping will round-off the driving effect of the SET and we find,
\begin{eqnarray*}
\delta x^2&\simeq&\delta x^2_{i}\frac{\gamma_{i}}{\gamma_{e}}\\
\delta u^2&\simeq&\delta u^2_{i}\frac{\gamma_{i}}{\gamma_{e}}
\end{eqnarray*}
and
\begin{equation}
k_{\rm B}T_{\rm eff}\simeq \frac{\gamma_{i}}{\gamma_{\rm
e}}eV_{\rm ds}\langle P\rangle_{N}\langle P\rangle_{N+1}. \label{teff}
\end{equation}
These expressions are consistent with the average back-action
displacement noise derived in Ref.\ [\onlinecite{ZB}].

The relatives value of the intrinsic and extrinsic damping
constants are also important for understanding why a single-mode
description of the resonator is sufficient. It turns out that the
magnitude of the intrinsic damping constant is very much larger
for the fundamental flexural mode of a resonator than for higher
modes. It is for this reason that the behavior of the system is
dominated by the fundamental mode and our single-mode treatment is
appropriate, as we discuss in appendix C.

\subsection{Finite Oscillator Temperature}
 The electrons
in the SET and the resonator will be in contact with environmental
degrees of freedom which will act as thermal reservoirs at finite
temperature, but for the mesoscopic system we are considering,
these temperatures will not necessarily be the same and their
effect on the dynamics may be very different. The effect of the
finite electron temperature, $T_{e}$, will be to round out the
tunnelling rates and thereby change the charge dynamics of the SET
somewhat. However, from the full expression for electron
tunnelling rate, Eq.\ (\ref{ft}), we can see that  this effect
will be small so long as $k_{\rm B}T_{e}\ll eV_{\rm ds}$.

The effect of a non-zero external bath temperature for the
resonator, $T_{r}$, can be gauged by adding in one further
term to the coupled master equations, (\ref{spn}) and
(\ref{spn1}), describing diffusion in velocity space. Thus in this
case the extra terms  take the form,
\[
\frac{\partial }{\partial u}\left(\gamma_e u P_{N(N+1)}
\right)+\frac{\partial}{\partial u}\left(
\frac{G}{2m^2}\frac{\partial P_{N(N+1)}}{\partial u}\right),
\]
so that without the coupling to the SET, the equation of motion
for the resonator probability distribution is simply Kramer's
equation.\cite{lemon} For systems close to equilibrium, $G$ is a
constant related to the background temperature by
\begin{equation}
G=2mk_{\rm B}T_{r}\gamma_{e}.\label{fd}
\end{equation}

In the regime where $\alpha,\epsilon,\kappa\ll 1$ we can
approximate the full equations for the variances by two the
expressions
\begin{eqnarray}
\delta x^2&\simeq& \frac{1}{m\omega_0^2}\left[ \frac{eV_{\rm
ds}\langle P\rangle_{N}\langle P\rangle_{N+1}}{1+\gamma_{e}/\gamma_{i}}
+\frac{k_{\rm B} T_{r}}{1+\gamma_{i}/\gamma_{e}}\right] \label{deltaxf}\\
 \delta u^2&\simeq&\frac{1}{m}\left[ \frac{eV_{\rm
ds}\langle P\rangle_{N}\langle P\rangle_{N+1}(1-\kappa)}{1+\gamma_{e}/\gamma_{i}}
+\frac{k_{\rm B} T_{r}}{1+\gamma_{i}/\gamma_{e}}\right]. \label{deltauf}
\end{eqnarray}
The derivation of these relations rests on the use of the
fluctuation-dissipation theorem [Eq.\ (\ref{fd})], whose validity
for the non-equilibrium SET-resonator system we have assumed
without proof. Nevertheless, the resulting expressions for the
variances have a very intuitive form. For $\kappa\ll 1$, the
effective oscillator temperature satisfies
\begin{equation}
    \frac{T_{\rm eff}}{Q_{\rm eff}}=\frac{T_{r}}{Q_{e}}+\frac{T_{i}}{Q_{i}},
    \label{balancetemp}
\end{equation}
where the extrinsic, intrinsic, and effective quality factors are
defined as $Q_{e}=\omega_{0}/\gamma_{e}$,
$Q_{i}=\omega_{0}/\gamma_{i}=1/(\kappa\epsilon)$, and $Q_{\rm
eff}^{-1}=Q_{e}^{-1}+Q_{i}^{-1}$, respectively. The intrinsic
temperature, $T_{i}$, is given by Eq.~(\ref{te}).
Eq.~(\ref{balancetemp}) is just the steady state relation for a
system in thermal contact with two reservoirs at temperatures
$T_{r}$ and $T_{i}$, and with the thermal conductivities of the
pathways between the system and each reservoir proportional to the
respective damping rates (and hence inversely proportional to the
quality factors).\cite{hirakawa}

The relative importance of
heating due to the SET and that due to the resonator's environment is
controlled by the ratios $Q_{i}/Q_{e}$ and $T_{i}/T_{e}$. When
$Q_{i}\ll Q_{e}$, we have $T_{\rm eff}=T_{i}+T_{r}Q_{i}/Q_{e}$.
As an example application, note that this equation predicts that it
would be possible to actually cool the resonator to a temperature
below $T_{r}$ by coupling it to an appropriate SET such that
$T_{i}<T_{r}$.\cite{hirakawa}
If, in addition to $Q_{i}\ll Q_{e}$, we  have that $T_{i}\gg T_{r}Q_{i}/Q_{e}$,
then the resonator hardly feels the
effect of the external environment and its state is determined by
the interaction with the SET electrons. In contrast, when
$Q_{i}\gg Q_{e}$, we recover Eq.~(\ref{teff}) provided
$T_{r}\ll T_{i}Q_{e}/Q_{i}$. If instead $T_{r}\gg T_{i}Q_{e}/Q_{i}$,
then the external environment
dominates over the SET electrons and effectively determines the
state of the resonator.

In the regime where $Q_{i}\gg Q_{e}$, but where the ratio
$T_{i}/T_{e}$ is arbitrary,  it is possible to
describe the dynamics of the SET-resonator system using a very
compact and intuitive formalism. Details of such an approach,
which compliments that given here, will be described
 elsewhere.\cite{yong}

\section{Current Characteristics of the SET}
The analysis we have performed so far describes the effects of the
electrons on the dynamics of the resonator. However, the electrons
passing through the SET are of course influenced by the motion of
the resonator and this is manifested as a change in the average
current through the SET.

In the absence of extrinsic damping, and at $T_{r}=T_{
e}=0$, the ensemble-averaged current through the SET can be
calculated from the tunnelling rates and the steady-state
probability distribution function:
\begin{eqnarray}
\langle I\rangle_{i}&=& e\int dx \int du\: \frac{\Delta E_L^+(x)}{Re^2}P_{N+1}(x,u)\\
&=& e\int dx \int du\:\frac{\Delta E_R^+(x)}{Re^2}P_{N}(x,u)\\
&=& \frac{e}{\tau_t}\langle P\rangle_{N}\langle P\rangle_{N+1}
\left( 1-\kappa\right)\\
&=& \frac{m}{Re^2}\delta u^2_{i}.
\end{eqnarray}
Comparing this with the fixed gate result,  Eq.\ (\ref{currfix}),
we see that the resonator slightly reduces the current flowing
through the SET.

Including extrinsic damping, we find that the averaged current
becomes
\begin{equation}
{\langle I\rangle}=\langle I\rangle_{
i}\left(\frac{\epsilon^2(\alpha+\kappa)
+\alpha(1+\alpha)}{\epsilon^2(\alpha+\kappa)
+\alpha(1+\alpha)(1-\kappa)}\right).
\end{equation}
The average current is modified by external damping because it
depends on the tunnelling rates through the individual junctions
which themselves depend on the properties of the sub-ensembles
through $\langle x \rangle_N$ and $\langle x \rangle_{N+1}$. The
external damping suppresses the motion of the resonator and this
is reflected in the fact that for $\alpha\gg \epsilon^2,
\kappa\epsilon^2$ the average current is effectively reduced to
that for the fixed-gate SET.

\section{Practical Considerations}

The predicted effect of the classical back-action of the SET on
the resonator  using currently available technology depends
crucially on the range of values of the parameters in our model
which are accessible in practice. Typical, achievable values for
the individual SET and resonator parameters are discussed in
detail in appendix D.

The maximum effective temperature of the resonator coupled to the
SET, neglecting the effect of the extrinsic damping, is  $eV_{\rm
ds}/4k_{\rm B}$. Considering, for example, a typical total SET capacitance
$C_{\Sigma}=0.5~{\rm fF}$ and drain-source voltage $V_{\rm
ds}=0.5e/C_{\Sigma}=0.16~{\rm mV}$, the maximum effective temperature
is about $0.5~{\rm K}$.  Expressed in terms of the various parameters, the
dimensionless coupling strength is
\begin{equation}
    \kappa=\left(\frac{C_{g}}{C_{\Sigma}}\right)^{2}\frac{V_{g}}{V_{\rm
    ds}}\frac{eV_{g}}{m\omega_{0}^{2}d^{2}}
    \label{kappaformula}.
\end{equation}
Substituting in the expressions for $C_{g}$, $m$, and $\omega_{0}$
appropriate for, say, Si and the above values for $C_{\Sigma}$ and $V_{\rm ds}$,
Eq.~(\ref{kappaformula}) becomes
\begin{equation}
    \kappa=1\times 10^{-12}V_{g}^{2}\frac{l^{5}w}{d^{4}t^{3}},
    \label{kappaform2}
\end{equation}
where the cantilever dimensions $l$, $w$, $t$, and cantilever-island
gap $d$ are expressed in micrometers, and the gate voltage $V_{g}$ is
given in Volts. Similarly, the intrinsic
quality factor is
\begin{equation}
    Q_{i}=1.3\times 10^{12}
    \frac{1}{V_{g}^{2}}\frac{t^{2}d^{4}}{l^{3}w},
    \label{qualityform}
\end{equation}
where again the dimensions are expressed in micrometers. Note the
strong dependence on the cantilever length $l$ and gap $d$. As might
be expected, decreasing the gap and increasing the length strengthens
the coupling between the SET and resonator and reduces the quality
factor. This trend can be clearly seen for the range of cantilever
length examples considered in Table~\ref{table1}. In each of the examples, the
transverse cantilever dimensions are $w=0.25~\mu{\rm m}$ and
$t=0.2~\mu{\rm m}$, and the gap is $d=0.1~\mu{\rm m}$.
\begin{table}
\caption{The resonant frequency, coupling strength and intrinsic
quality factor for Si cantilevers with different lengths.}
\label{Table 1}
\begin{tabular}{|c|c|c|c|}
\hline
$l$ in $\mu{\rm m}$ & $1$ & $5$ & $10$\\
\hline
$f_{0}$ in MHz & $240$ & $9.6$ & $2.4$\\
\hline
$\kappa/V_{g}^{2}$ in V$^{-2}$  & $3\times 10^{-7}$ & $9.8\times 10^{-4}$ & $0.03$\\
\hline
$Q_{i}V_{g}^{2}$ in V$^{2}$ & $2.1\times 10^{7}$ &
$1.7\times 10^{5}$ & $2.1\times 10^{4}$  \\
\hline
\end{tabular}
\label{table1}
\end{table}

For micron-sized resonators, the extrinsic quality factors,
$Q_{\rm e}=\omega_0/\gamma_{e}$, of the fundamental modes
typically lie in the range\cite{roukeshilton} $\sim 10^3$--$10^4$,
which, from the first column in Table~\ref{table1}, will be much
less than the
 quality factors associated with intrinsic damping even for gate
 voltages close to vacuum breakdown $\sim 10~{\rm V}$ (see appendix E). In this
regime, for the one micron long example the effective temperature will be given,
to a good approximation, by $T_{\rm eff}=T_{r}+T_{i}Q_{e}/Q_{i}<T_{r}+
0.24 V_{g}^{2}~{\rm mK}$.
 Therefore, the  maximum amount the SET can add to the cantilever
 temperature is about $24~{\rm mK}$ for $V_{g}$ tuned to current peak
 maximum around the breakdown voltage $=10~{\rm V}$. The description of the
 SET back-action  on the resonator in this regime coincides with the
 analysis given in Ref.~\onlinecite{ZB}. On the other
 hand,  increasing the cantilever length by just a factor of ten,
it is not difficult to be in the opposite, SET back-action
dominated regime.
 With, e.g., $V_{g}=5~{\rm V}$ and
 $Q_{e}=10^{4}$, we have $Q_{e}/Q_{i}=12$ and
 $T_{\rm eff}=T_{i}+0.084 T_{r}\simeq T_{i}$ for $T_{r}\ll 12 T_{i}$.

\section{Conclusions and discussion}
We have carried out a detailed analysis of the dynamics of a
nano-mechanical resonator coupled to a SET. For the isolated
SET-resonator system, the electrons tunnelling through the SET act
like an effective thermal bath for the resonator. The variances of
the stationary probability distribution for the resonator, and the
dynamics associated with its approach to the steady-state, allow
us to assign an effective temperature and intrinsic damping
constant due to the electrons, so long as the coupling between the
resonator and the SET is small, i.e., $\kappa\ll 1$.

The intrinsic damping constant controls the time taken for the
isolated system to reach a steady-state. The effective temperature
for the isolated SET-resonator system does not depend on the
coupling between the resonator and the SET (for fixed values of
$\langle P \rangle_{N}$ and $\langle P\rangle_{N+1}$), a fact
which is somewhat surprising at first sight. However, although the
average energy of the resonator in the steady-state is independent
of the coupling, $\kappa$, the time taken for the energy stored in
the resonator to reach this average, constant level {\emph is}
determined by the coupling. In the limit that the SET-resonator
coupling goes to zero, the time taken to reach a steady-state will
diverge and so the whole concept of a steady-state for the coupled
system and hence an effective temperature breaks down. The effects
of extrinsic damping can be understood as trying to take energy
out of the resonator, through an additional pathway, on a
time-scale set by the extrinsic damping rate. Hence, the effective
temperature of the resonator will now depend on the relative
magnitudes of the extrinsic and intrinsic damping constants.

 We can also draw some conclusions from
our results for the coupled dynamics which are more widely
relevant to the study of nano-electromechanical systems. The idea
that a mechanical resonator coupled to a mesoscopic conductor
should have an intrinsic damping rate was demonstrated by Mozyrsky
and Martin\cite{noise} and confirmed by Smirnov {\it et
al.}\cite{noise2} The results we obtain for a classical resonator
demonstrate the generality of this result.  Thus far, many studies
of mesoscopic conductors coupled to nano-mechanical
resonators\cite{cs,qs,brandes} do not discuss the possibility of
damping of the resonator arising from its interaction with the
electrons. Indeed, in many nano-electromechanical systems it may
be that there is a non-zero intrinsic damping rate which affects
the character of the steady-state.

In this work, we have concentrated almost exclusively on how the
properties of a nano-mechanical resonator are affected by coupling
to a SET. Although we have calculated the effect of the coupling
to the resonator on the average current through the SET, the
spectrum of the current-noise has not yet been obtained, though
work on this is in progress.\cite{cnoise} The current-noise
spectrum will give important further information about the coupled
dynamics of the system as it is very sensitive to correlations
between the charge on the SET island and the position of the
resonator.

While preparing this work we became aware of Ref.\
[\onlinecite{mmh}] which considers the quantum dynamics of the
SET-resonator system, but in a different regime to that considered
here.

\section*{Acknowledgements}
We would like to thank O. Buu, A. Korotkov,
A. Martin, R. Onofrio, S. A. Ramakrishna,  K.
Schwab, and M. Wybourne for a number of very useful discussions. MPB and YZ
were supported in part by an award from the Research Corporation, and
by the National Security Agency (NSA) and Advanced Research and
Development Activity (ARDA) under Army Research Office (ARO)
Contract No.\ DAAG190110696.

\appendix
\section{Total energy and tunnelling rates}
The circuit diagram for a nano-mechanical resonator coupled
capacitively to the island of a SET is shown schematically in
Fig.\ \ref{fig:schema}, where a nano-mechanical resonator replaces
the usual fixed voltage gate for the SET. In order to describe the
dynamics of the coupled SET-resonator system, we require
expressions for the total energy with $N$ and $N+1$ electrons on
the SET island as well as for the tunnelling rates through the two
junctions.

\subsection{Total energy}
The total energy of the coupled SET-resonator system for a given
number of charges on the island is most easily obtained by
generalizing the usual expression for the energy of charges on an
SET island with a fixed voltage gate,\cite{Makhlin} given in Sec.\
II. Here we treat the resonator as a single-mode harmonic
oscillator. The generalization to several modes is discussed in
appendix C.

When the fixed gate is replaced by a harmonic oscillator, the
gate capacitance, and hence the energy of the system, depends on
the position of the oscillator. If the distance between the
unperturbed oscillator and the SET island is $d$ and the
displacement, $q$, of the oscillator is always much less than this
distance then the gate capacitance can be approximated by,
\begin{equation}
C_g(q)=C^0_g(1-q/d)
\end{equation}
where $C_g^0$ is the capacitance of the un-displaced oscillator.
Strictly-speaking we should write the approximation as
$C_g(q)=C^0_g+q_0(dC_g/dq_0)_{q_0=0}$ where $q_0$ is the normal
mode displacement of the fundamental mode, but we will neglect
this detail as it leads to an unimportant geometrical correction.

For nano-mechanical resonators, the values of $C_g^0$ are
typically much less than the SET junction capacitances $C_j$.
Hence, we will neglect the position dependence of $C_{\Sigma}$ and
 deal only with position dependence in the term $-2NN_g$. Thus,
the charging energy of the SET island with $N$ excess electrons
can be written as,
\begin{equation}
E_{ch}=E_c\left(N^2-2NN_g^0+2NN_g^0\frac{q}{d}\right),
\end{equation}
where $E_c=e^2/(2C_{\Sigma})$ and $N_g^0=-C_g^0 V_g/e$ (for
simplicity we will drop the superscripts from now on as $N_g$ and
$C_g$ will always take this form in what follows).

The polarization charge induced by the gate can make the charging
energies of states with $N$ and $N+1$ electrons on the SET island
degenerate. For convenience, we define
\begin{equation}
\delta N=N_g-N-\frac{1}{2}.
\end{equation}
Using this notation, we find
\begin{eqnarray}
N^2-2NN_g&=&\delta N^2+\frac{1}{4}+\delta N-N_g^2\\
(N+1)^2-2(N+1)N_g&=&\delta N^2+\frac{1}{4}-\delta N-N_g^2.
\end{eqnarray}
Thus, we can write the total energy of the coupled SET-resonator
system with $N$ and $N+1$ excess electrons as,
\begin{eqnarray}
H_N&=&E_c\delta N+2E_cN_g\frac{q}{d}\left(N_g-\frac{1}{2}-\delta N\right)
+\frac{p^2}{2m}+\frac{1}{2}m\omega_0^2q^2\\
H_{N+1}&=&-E_c\delta
N+2E_cN_g\frac{q}{d}\left(N_g+\frac{1}{2}-\delta N\right)
+\frac{p^2}{2m}+\frac{1}{2}m\omega_0^2q^2,
\end{eqnarray}
respectively, where we have dropped terms which are constant in
both.  Making the change of coordinates $x=q+\Delta x$ to
eliminate the linear term in $H_N$, we have
\begin{equation}
\frac{1}{2}m\omega_0^2x^2=\frac{1}{2}m\omega_0^2q^2+2E_cN_g\frac{q}{d}
\left(N_g-\delta
N-\frac{1}{2}\right)+\frac{1}{2}m\omega_0^2\Delta x^2,
\end{equation}
where
\begin{equation}
\Delta x=\frac{2E_cN_g}{m\omega_0^2 d}\left(N_g-\delta
N-\frac{1}{2}\right).
\end{equation}

Expressing the total energies of the coupled system with $N$ and
$N+1$ charges in terms of the new variable $x$, we obtain:
\begin{eqnarray}
H_N&=&E_c\delta N +\frac{p^2}{2m}+\frac{1}{2}m\omega_0^2x^2\\
H_{N+1}&=&-E_c\delta N
+\frac{p^2}{2m}+\frac{1}{2}m\omega_0^2(x-x_0)^2-\frac{1}{2}m\omega_0^2x_0^2\\
\nonumber
 &&-m\omega_0^2x^2_0(N_g-\delta N -1/2 ),
\end{eqnarray}
where $x_0=-2E_cN_g/(m\omega_0^2 d)$ and we have again dropped
constant terms.
\subsection{Tunnelling rates}
Within the orthodox model of single electron tunnelling, we can
derive simple expressions for the tunnelling rates in either
direction through the two SET junctions. Adopting the convention
that electron tunnelling forwards corresponds to leftwards motion
in Fig.\ \ref{fig:schema}, we can write down expressions for the
four possible free energy differences:
\begin{eqnarray}
\Delta E^+_L&=&-\Delta E^-_L=H_{N+1}-H_N-\mu_L\\
\Delta E^+_R&=&-\Delta E^-_R=H_{N}-H_{N+1}+\mu_R,
\end{eqnarray}
where $\mu_{L(R)}$ is the chemical potential of the
left(right)-hand lead. Using the expressions for $H_{N}$ and
$H_{N+1}$ obtained above we find,
\begin{eqnarray}
\Delta E^+_L&=& -2E_c\delta
N-m\omega_0^2x_0x-m\omega_0^2x^2_0(N_g-\delta N -1/2 )+\frac{eV_{\rm ds}}{2}\\
\Delta E^+_R&=& 2E_c\delta
N+m\omega_0^2x_0x+m\omega_0^2x^2_0(N_g-\delta N -1/2
)+\frac{eV_{\rm ds}}{2},
\end{eqnarray}
where we have used the fact that $-\mu_L=\mu_R=eV_{\rm ds}/2$.

\section{Numerical integration of the probability density}

In this appendix we outline the way in which the coupled master
equations, (\ref{pn}) and (\ref{pn1}), can be integrated
numerically. The coupled equations of motion for the probability
distributions contain terms describing the free evolution of the
resonator, which are deterministic, and terms describing the
tunnelling of electrons which are essentially stochastic. The free
evolution of the resonator is known exactly and the problem of
obtaining the dynamics of the overall distribution  is simplified
considerably be disentangling this part of the motion from the
electron tunnelling by transforming to an interaction picture.

Formally, we can write the Liouville equation\cite{Pg} for
$P_N(x,u;t)$,
\begin{equation}
i\frac{dP_N(x,u;t)}{dt}=L_N P_N(x,u;t),
\end{equation}
where $L_N$ is the Liouville operator and  a very similar equation
for $P_{N+1}(x,u;t)$. The dynamics of the resonator  is simply
that of a harmonic oscillator, albeit one whose equilibrium
position shifts by $x_0$ when an extra electron joins the island,
together with a perturbation arising from tunnelling of electrons
onto and off the island. Therefore, we can simplify the problem
considerably by transforming into an interaction picture. We split
the Liouville operator into two parts,
\begin{equation}
L_N=L_N^0+\delta L,
\end{equation}
where $L_N^0P_N(x,u;t)=i\left\{H_N,P_N(x,u;t)\right\}$ describes
the free evolution of the oscillator with $N$ charges on the
island (for $N+1$ charges, we define
$L_{N+1}^0P_{N+1}(x,u;t)=i\left\{H_{N+1},P_{N+1}(x,u;t)\right\}$
by analogy) and $\delta L$ describes the electron tunnelling
processes. Defining  $\tilde{P}_N(t)={\rm e}^{+iL^0_Nt}P_N(t)$,
and a similar relation for $N+1$, we have for the equations of
motion in the interaction picture
\begin{eqnarray}
\frac{d{\tilde P}_N(x,u;t)}{dt}&=&\frac{1}{Re^2}
\left\{ \Theta[\Delta E_L^{+}(x')]\Delta E_L^{+}(x')+
\Theta[\Delta E_R^{-}(x')]\Delta E_R^{-}(x')\right\}\\
&&\times{\tilde P}_{N+1}(x+x_0(1-\cos(\omega_0 t)),u+
\omega_0 x_0\sin(\omega_0 t);t) \nonumber\\
&&-\frac{1}{Re^2}\left\{ \Theta[\Delta E_R^{+}(x')]
\Delta E_{R}^{+}(x')+\Theta[\Delta E_L^{-}(x')]\Delta E_L^{-}(x')\right\}\\
 &&\times{\tilde P}_N(x,u;t)\nonumber \\
\frac{d{\tilde P}_{N+1}(x,u;t)}{dt}&=&-\frac{1}{Re^2}
\left\{ \Theta[\Delta E_R^{+}(x'')]\Delta E_{R}^{+}(x'')+
\Theta[\Delta E_L^{-}(x'')]\Delta E_L^{-}(x'')\right\}\\
&&\times {\tilde P}_{N+1}(x,u;t)\nonumber \\
&&+\frac{1}{Re^2} \left\{ \Theta[\Delta E_L^{+}(x'')]\Delta
E_L^{+}(x'')+
\Theta[\Delta E_R^{-}(x'')]\Delta E_R^{-}(x'')\right\}\\
&&\times{\tilde P}_N(x-x_0(1-\cos(\omega_0 t)),u-\omega_0
x_0\sin(\omega_0 t);t)\nonumber\\
\end{eqnarray}
where
\begin{eqnarray}
x'&=&\left[x\cos(\omega_0 t)-\frac{u}{\omega_0} \sin(\omega_0
t)\right]\\
x''&=&\left[x_0+(x-x_0)\cos(\omega_0 t)-\frac{u}{\omega_0}
\sin(\omega_0 t)\right].
\end{eqnarray}
Thus the transformation to the interaction picture eliminates the
free-evolution of the oscillator and leaves us with a pair of
simple rate equations which can be integrated numerically in a
straightforward way.

Integrating these equations numerically on a finite grid and then
performing the reverse transformation back from the interaction
picture to obtain the full time dependence, we can calculate the
probability distribution as a function of time. The accuracy of
the calculation is set primarily  by the size of the grid used,
but convergence of the results can be confirmed by performing the
calculation on grids with progressively smaller mesh sizes.
\section{Higher modes of the resonator}

In the main body of this paper we treat the resonator as a
single-mode harmonic oscillator and neglect the effect of all but
the fundamental mode. In this appendix we show why such an
approximate description works well.

In most studies of nano-electromechanical systems in which the
mechanical component consists of a beam or cantilever, which has a
spectrum of vibrational modes, all but the fundamental flexural
mode is usually neglected. For systems where the coupling between
modes is weak, this is a sensible approximation as it turns out
that geometrical factors mean that the coupling between a normal
mode of a beam or cantilever decreases rapidly with increasing
frequency.\cite{wp}

In the system we consider here, the situation is a little
different as the electrons passing through the SET act as a
thermal bath and heat the resonator up to an effective temperature
that is independent of the coupling. Therefore, in the absence of
extrinsic damping each of the normal modes would have the same
effective temperature. The overall variance in the position of the
resonator would then be obtained by summing over the contribution
from each of the normal modes. However, when we include the effect
of external damping (discussed in Sec.\ IV,) it is found that the
effective temperature for a single mode depends on the ratio of
intrinsic to extrinsic damping rates: $\gamma_{i}/\gamma_{e}$.
When the ratio $\gamma_{i}/\gamma_{e}$ is $\ll 1$, the extrinsic
damping dominates and the SET has little effect on the mode.
However, for $\gamma_{i}/\gamma_{e}\leq 1$ the extrinsic damping
does not fully counteract the thermal-like excitation of the mode
due to the SET.

The intrinsic damping, $\gamma_{i}$, can be written in terms of
the basic parameters of the system as
\begin{equation}
\gamma_{\rm_i}=\frac{Re^3}{m d^2}\left(\frac{C_gV_g}{C_{\Sigma}
V_{\rm ds}}\right)^2.
\end{equation}
This formula follows from the approximation we made in appendix A
where we neglected the distinction between the position of the
center of mass of the resonator, $q$, and the normal-mode
displacement of the fundamental mode. In fact, the linear
correction to the gate capacitance should be written as
$(dC_g/dq_0)_{q_0=0}=-\chi C_g/d$, where $q_0$ is the fundamental
mode position co-ordinate and $\chi$ is a geometrical correction
factor which we have neglected as it is of order unity for the
fundamental mode.

In general, there will be an intrinsic damping constant for each
normal mode of the resonator, $n$, of the form
\begin{equation}
\gamma_{\rm_i}^n=\frac{Re^3}{m }\left(\frac{V_g}{C_{\Sigma}V_{\rm
ds}}\right)\left(\frac{dC_g}{dq_n}\right)_{q_n=0}^2,
\end{equation}
where $q_n$ is the displacement of the $n$-th normal
mode.\cite{modes,wp} However, the intrinsic damping  constant for
higher modes will always be {\it much less} than that of the
fundamental mode because of their shorter wavelengths and the
presence of nodes, which significantly reduce the value of the
coupling, given by $(dC_g/dq_n)_{q_n=0}$.\cite{modes} Furthermore,
there is reason to expect that the extrinsic damping,
$\gamma_{e}^n$, will be greater for higher-frequency
modes,\cite{Rast} leading to a further reduction in the ratio
$\gamma_i^n/\gamma_e^n$, and hence the effect of these modes on
the overall motion of the resonator.

The current characteristics of the SET and the associated noise
spectrum are also much more strongly affected by the fundamental
mode of the resonator than by higher modes. This is again because
of the weak-coupling between the resonator and the SET.

It should be noted that the treatment here could be extended to
include the effect of higher modes of the resonator. The
calculations presented in the main text consider a single mode,
and the correct multi-mode results could easily be obtained by
adding up the contributions of each of the modes, although the
appropriate coupling constant, $\kappa$ (which in general would
depend on the frequency and geometry of the mode), would need to
be used for each of the modes. The only assumption in our approach
is that the normal modes of the resonator are effectively
uncoupled from each other on all the time-scales of interest.

\section{Parameter Values}
In this appendix we discuss the physical values of the various
parameters in our model of the SET-resonator system which could be
achieved  using available technology. We concentrate
on the parameter values which would be necessary for the system to
operate in the regime we have explored theoretically, where the
SET is well-described by the orthodox model, and the drain-source voltage is
large enough that the resonator behaves classically.

For the orthodox model to apply, the
charging energy, $E_c=e^2/(2C_{\Sigma})$, must dominate over the
electron thermal energy scale, $k_{\rm B}T_{e}$. The total capacitance of the
SET, $C_{\Sigma}$, is dominated by the capacitance of the
tunnel junctions, $C_j$,  typically  much larger than that of the
mechanical gate, $C_g$. A typical choice of  $C_j= 0.25$~fF
gives $E_c/k_{\rm B}\simeq 2$~K so that at electron temperatures of a few
tens of mK, the charging energy clearly dominates. The junction
resistances, $R$, of the SET must be set relatively high, e.g., $R=
100~$k$\Omega$, in order to suppress higher-order electron
tunnelling processes which cannot be described within the orthodox
model. The drain-source voltage  is typically chosen to be
of order the charging energy $eV_{\rm ds}\sim e^2/(2C_{\Sigma})$, as
it is in this regime that the SET conductance is most sensitive to
changes in the gate capacitance. Thus for $C_j= 0.25$fF, we
have, e.g., $V_{\rm ds}=e/(2C_{\Sigma})=0.16$~mV and the electron tunnelling
time, $\tau_t$, is consequently $2RC_{\Sigma}= 10^{-10}$~s.

The properties of the nano-mechanical resonator used as a
voltage-gate can be varied to quite a high degree. The frequency
of a nano-mechanical resonator is determined by the material it is
made from and its geometrical form. For a cantilever, a beam which
is clamped at one end and free at the other, the fundamental
flexural mode frequency is given by $\omega_0=1.02
(t/l^2)\sqrt{E/\rho}$, where $l$ is the length, $t$ the thickness,
$E$ is Young's modulus, and $\rho$ the mass density of the
material from which the resonator is fabricated. Resonators
fabricated from semiconductors typically have $\sqrt{E/\rho}$ in
the range\cite{sic} $\sim 10^3$--$10^4$~ms$^{-1}$. For the example
of a Si cantilever with $E=1.33\times 10^{11}~{\rm Nm}^{-2}$ and
$\rho=2.33\times 10^{3}~{\rm kg m}^{-3}$, the fundamental flexural
mode frequency is $f_{0}=1.2 t/l^{2}~{\rm GHz}$, with the
dimensions expressed in micrometers. The gate capacitance of a
nano-mechanical resonator is also determined by geometry. For a
vacuum gap, $C_{g}=\lambda\epsilon_{0}lw/d$, where $w$ is the
cantilever width, $d$ the cantilever-SET island gap, and $\lambda$
is a geometrical factor (which  accounts, e.g., for the fact that
the metal electrode may have smaller area than the cantilever
surface $lw$). Considering, e.g., a resonator with dimensions
$l=1~\mu$m,  $w=0.25~\mu$m, $t=0.2~\mu$m, $d=0.1~\mu$m, and
$\lambda=1/3$,  we have $C_g= 74$~aF and $f_{0}=244~{\rm MHz}$.
This corresponds to $\epsilon=0.02$ for $\tau_t=10^{-10}$~s. The
resonator would be well within the classical regime as the energy
quanta for this frequency are $\hbar\omega_0=0.15\mu$eV, very much
less than the energy associated with the above drain-source
voltage.

The magnitude of the gate voltage,
$V_g$, is limited by the need to avoid electrostatic breakdown in
the dielectric between the SET and the resonator;\cite{ABS,bw,schwab}
for vacuum the breakdown voltage is about $10^{8}~{\rm V}/{\rm
m}$,\cite{ma} translating into a maximum voltage of about $10~{\rm V}$
for a resonator-SET island separation $d=0.1~\mu{\rm m}$.


\begin{thebibliography}{99}
\bibitem{Henry} X.M.H. Huang, C.A. Zorman, M. Mehragany, and M.L.
Roukes, Nature (London) {\bf 421}, 496 (2003).
\bibitem{roukes} M. Roukes, { Physics World} {\bf 14}, 25
(2001).
\bibitem{blickrev} R.H. Blick, A. Erbe, L. Pescini, A. Kraus, D.V.
Scheible, F.W. Beil, E. Hoehberger, A. Hoerner, J. Kirschbaum, H.
Lorenz, and J.P. Kotthaus, J. Phys. Cond. Matt. {\bf 14}, R905
(2002).
\bibitem{elect} A.N. Cleland and M.L. Roukes, Nature (London) {\bf 392}, 160
(1998).
\bibitem{QD}J. Kirschbaum, E.M. H\"{o}hberger,  R.H. Blick, W. Wegscheider,
and M. Bichler, Appl. Phys. Lett. {\bf 81}, 280 (2002).
\bibitem{QPC}A.N. Cleland, J.S. Aldridge, D.C. Driscoll, and A.C. Gossard,
Appl. Phys. Lett. {\bf 81}, 1699 (2002).
\bibitem{set}  R.S. Knobel and A.N. Cleland,  Nature (London) {\bf 424}, 291 (2003).
\bibitem{cs}L.Y. Gorelik, A. Isacsson, M.V. Voinova, B. Kasemo,
R.I. Shekhter, and M. Jonson, Phys. Rev. Lett. {\bf 80}, 4526.
(1998); A. Isacsson, L.Y. Gorelik, M. V. Voinova, B. Kasemo, R.I.
Shekhter, and M. Jonson, Physica B {\bf 255}, 150 (1998); T. Nord,
L.Y. Gorelik, R.I. Shekhter, and M. Jonson, Phys. Rev. B {\bf 65},
165312 (2002).
\bibitem{shutrev} R.I. Shekhter, Yu. Galperin, L.Y. Gorelik, A.
Isacsson, and M. Jonson, J. Phys. Cond. Matt. {\bf 15}, R441
(2003).
\bibitem{qs}A.D. Armour and A. MacKinnon, Phys. Rev. B {\bf 66},
035333 (2002).
\bibitem{nish} N. Nishiguchi, Phys. Rev. B {\bf 65}, 035403
(2001).
\bibitem{nonlin}A. Erbe, H. Kr\"ommer, A. Kraus, R.H. Blick, G.
Corso, and K. Richter, App. Phys. Lett. {\bf 77}, 3102 (2000).
\bibitem{ron}R. Lifshitz and M.C. Cross, Phys. Rev. B {\bf 67}, 134302 (2003).
\bibitem{ABS}A.D. Armour, M.P. Blencowe, and K. C. Schwab, Phys.
Rev. Lett. {\bf 88}, 148301 (2002).
\bibitem{wp}A.D. Armour and M.P. Blencowe, Phys. Rev. B {\bf 64},
035311 (2002).
\bibitem{squeeze} M.P. Blencowe and M.N. Wybourne, Physica B {\bf
280}, 555 (2000).
\bibitem{noise}D. Mozyrsky and I. Martin, Phys. Rev. Lett. {\bf 89}, 018301
(2002).
\bibitem{white} J.D. White, Jap. J. Appl. Phys. Part 2 {\bf 32},
l1571-L1573 (1993).
\bibitem{bw}M.P. Blencowe and M.N. Wybourne, App. Phys. Lett. {\bf 77}, 3845
(2000).
\bibitem{ZB} Y. Zhang and M.P. Blencowe, J. Appl. Phys. {\bf 91},
4249 (2002).
\bibitem{BK}V. B. Braginsky and F. Ya. Khalili, {\em Quantum Measurement},
(Cambridge University Press, Cambridge, UK, 1992).
\bibitem{noise2}A.Yu. Smirnov, L.G. Mourokh, and N.J.M. Horing,
Phys. Rev. B {\bf 67}, 115312 (2003).
\bibitem{CL}A.O. Caldeira and A.J. Leggett, Ann. Phys. (NY) {\bf
149}, 374 (1983).
\bibitem{ult} Note, however, within the framework of our classical
 description of the SET, the electron temperature cannot be arbitrarily small
as discussed by H. Schoeller and G. Sch\"on, Phys. Rev. B {\bf
50}, 18436 (1994).
\bibitem{martinis}R.L. Kautz, G. Zimmerli, and J.M. Martinis,
J. Appl. Phys. {\bf 73}, 2386 (1993).
\bibitem{korotemp}A.N. Korotkov, M.R. Samuelsen, and S.A. Vasenko, J.
Appl. Phys. {\bf 76}, 3623 (1994).
\bibitem{set_nat}M.H. Devoret and R.J. Schoelkopf, Nature (London)
{\bf 406}, 1039 (2000).
\bibitem{rfset}R.J. Schoelkopf, P. Wahlgren, A.A. Delsing, and D.
Prober, Science {\bf 280}, 1238 (1998).
\bibitem{elect2}K.W. Lehnert, K. Bladh, L.F. Spietz, D.
Gunnarsson, D.I. Schuster, P. Delsing, and R.J. Schoelkopf, Phys.
Rev. Lett. {\bf 90}, 027002 (2003).
\bibitem{FG} D.K. Ferry and  S.M. Goodnick, {\em Transport in
Nanostructures}, (Cambridge University Press, Cambridge, UK,
1997).
\bibitem{Makhlin}A. Shnirman and G. Sch\"{o}n, Phys. Rev. B {\bf 57}, 15400
(1998); Y. Makhlin, G. Sch\"{o}n and A. Shnirman, Rev. Mod. Phys.
{\bf 73}, 357 (2001).
\bibitem{ford}M. Amman, R. Wilkins, E. Ben-Jacob, P.D. Maker, and
R.C. Jaklevic, Phys. Rev. B {\bf 43}, 1146 (1991).
\bibitem{lemon}D.S. Lemons, {\em An Introduction to Stochastic
Processes}, (John Hopkins University Press, Baltimore, MD, 2002).
\bibitem{clelbook}A.N. Cleland {\em Fundamentals of
Nanomechanics}, (Springer-Verlag, Heidelberg, Germany, 2002).
\bibitem{hirakawa}H. Hirakawa, S. Hiramatsu, and Y. Ogawa, Phys. Lett.
A {\bf 63} 199 (1977).
\bibitem{yong} Y. Zhang {\it et al.}, (in preparation).
\bibitem{roukeshilton}M.L. Roukes, Tech. Digest 2000 Solid-State
Sensor and Actuator Workshop, Transducer Research Foundation,
Cleveland, 2000; cond-mat/0008187.
\bibitem{brandes}T. Brandes and N. Lambert, Phys. Rev. B {\bf
67}, 125323 (2003).
\bibitem{cnoise} A.D. Armour and M.P. Blencowe, (in
preparation)
\bibitem{mmh} D. Mozyrsky, I. Martin, and M.B. Hastings, cond-mat/0306480.
\bibitem{Pg}I. Prigogine, {\em Non-Equilibrium Statistical
Mechanics}, (Interscience, John Wiley, NY, 1962).
\bibitem{modes}H.-J. Butt and M. Jaschke, Nanotechnology {\bf 6},
1 (1995).
\bibitem{Rast}S. Rast, C. Wattinger, U. Gysin, and E. Meyer, Nanotechnology
{\bf 11}, 169 (2000).
\bibitem{schwab}K.C. Schwab, Appl. Phys. Lett. {\bf 80}, 1276 (2002).
\bibitem{sic}Y.T. Yang, K.L. Ekinci, X.M.H. Huang, L.M. Schiavone, M.L.
Roukes, C.A. Zorman, and M. Mehragany,  Appl. Phys. Lett. {\bf
78}, 162 (2001).
\bibitem{ma}X. Ma and T.S. Sudarshan, J. Vac. Sci. Technol. B
{\bf 16}, 1174 (1998).
\end{thebibliography}
\end{document}